\documentclass[aps,pra,twocolumn,showpacs,nofootinbib]{revtex4}

\usepackage[english]{babel}
\usepackage{amsmath}
\usepackage{amsthm}
\usepackage{amssymb}
\usepackage{booktabs}
\usepackage{color}
\usepackage{hyperref}
\hypersetup{
  colorlinks   = true,  
  urlcolor     = blue,  
  linkcolor    = blue,  
  citecolor    = red    
}
\usepackage{natbib}
\usepackage{pgfplots}
\usepackage{subfigure}
\usepackage{url}

\newcommand{\refFigure}[1]{{\textrm{Fig.~\ref{#1}}}}
\newcommand{\refTable}[1]{{\textrm{Table~\ref{#1}}}}
\newcommand{\refEquation}[1]{{\textrm{Eq.~(\ref{#1})}}}

\newcommand{\refSection}[1]{{\textrm{\S\ref{#1}}}}
\newcommand{\refAppendixSection}[1]{{\textrm{Appendix \ref{#1}}}}

\newcommand*{\bj}{\skew{3}{\hat}{j}}  

\usepackage{scalerel,stackengine}
\stackMath
\newcommand\reallywidehat[1]{%
\savestack{\tmpbox}{\stretchto{%
  \scaleto{%
    \scalerel*[\widthof{\ensuremath{#1}}]{\kern-.6pt\bigwedge\kern-.6pt}%
    {\rule[-\textheight/2]{1ex}{\textheight}}
  }{\textheight}%
}{0.5ex}}%
\stackon[1pt]{#1}{\tmpbox}%
}
\usepackage{comment}
\usepackage{float}

\begin{document}

\title{Finite-range effects in Efimov physics beyond the separable approximation}

\author{P. M. A. \surname{Mestrom}}
\email{p.m.a.mestrom@tue.nl}

\author{T. \surname{Secker}}

\author{R. M. \surname{Kroeze}}
\altaffiliation[Current address: ]{Department of Physics, Stanford University, Stanford, California 94305, USA}

\author{S. J. J. M. F. \surname{Kokkelmans}}
\affiliation{Eindhoven University of Technology, P.~O.~Box 513, 5600 MB Eindhoven, The Netherlands}

\date{\today}


\begin{abstract}
We study Efimov physics for three identical bosons interacting via a pairwise square-well potential,
analyze the validity of the separable approximation as a function of the interaction strength, and investigate what is needed to improve  this approximation. We find separable approximations to be accurate for potentials with just one (nearly) bound dimer state. For potentials with more bound or almost bound dimer states, these states need to be included for an accurate determination of the Efimov spectrum and the corresponding three-body observables. We also show that a separable approximation is insufficient to accurately compute the trimer states for energies larger than the finite-range energy even when the two-body $T$ matrix is highly separable in this energy regime. Additionally, we have analyzed three distinct expansion methods for the full potential that give exact results and thus improve on the separable approximation. With these methods, we demonstrate the necessity to include higher partial-wave components of the off-shell two-body $T$ matrix in the three-body calculations. Moreover, we analyze the behavior of the Efimov states near the atom-dimer threshold and observe the formation of non-Efimovian trimer states as the potential depth is increased. Our results can help to elaborate simpler theoretical models that are capable of reproducing the correct three-body physics in atomic systems.
\end{abstract}

\maketitle

\section{Introduction}

Over the last decade, impressive experimental and theoretical progress has been made in the investigation of Efimov physics \cite{efimov1970energy, efimov1971weakly, braaten2006universality, naidon2017review}. Already in 1970, Efimov predicted that three particles interacting via short-range interactions exhibit an infinite sequence of three-body bound states scaling universally when the $s$-wave two-body scattering length $a$ diverges \cite{efimov1970energy,efimov1971weakly}. The Efimov effect has been predicted to occur universally over a range of different physical systems, such as nucleons \cite{jensen2004halos,hammer2010nucleons}, magnons \cite{nishida2013magnons} and atoms \cite{esry1999recombination}. Efimov physics was first observed experimentally in an ultracold atomic gas in 2006 \cite{kraemer2006cesium133}, by tuning the interaction strength using a Feshbach resonance. Since then, experiments in ultracold quantum gases have been successfully performed with different atomic species determining the three-body parameter $a_{-}$ \cite{kraemer2006cesium133, berninger2011cesium133, huang2014cesium133, zaccanti2009potassium39, pollack2009lithium7, dyke2013lithium7, gross2009lithium7, gross2010lithium7, gross2011lithium7, ottenstein2008lithium6, huckans2009lithium6, williams2009lithium6, wild2012rubidium85, barontini2009mixture, knoop2012helium4}, which is the scattering length where the ground Efimov state emerges from the three-body continuum.  These results combined gave rise to a species-independent value of $a_{-} \approx -9$~$r_{vdW}$ within $\pm 20 \%$ deviation \cite{naidon2017review}, where $r_{vdW}$ is the van der Waals length defined by $r_{vdW}~\equiv~  (m C_6/\hbar^2)^{1/4}/2$. Most importantly, it was demonstrated that the three-body parameter is fixed to the range of the interparticle interactions, which for the atomic trimers is related to the long-range behavior of van der Waals potentials via the dispersion coefficient $C_6$ which depends on the atomic species. The origin of this unexpected van der Waals universality of $a_{-}$ was successfully interpreted in recent theoretical works \cite{wang2012origin, dincao2015heteronuclear, schmidt2012universality, naidon2014physicalorigin, blume2015TBPhelium, langmack2018finiteAbg}, in which the dominance of the long-range behavior of the interaction potential on the three-body parameter was evidenced. 

Despite this success of van der Waals and finite-range models in general, there are still unexplained issues.
For example, the experimentally determined three-body parameters near narrow Feshbach resonances are in disagreement with theoretical expectations \cite{roy2013K39narrowFB} and also at positive scattering lengths several non-universal Efimov-related effects have been observed \cite{kraemer2006cesium133, berninger2011cesium133, ferlaino2011cesium133, knoop2009cesium133, zenesini2014cesium133, pollack2009lithium7, dyke2013lithium7, gross2009lithium7, gross2010lithium7, gross2011lithium7, machtey2012lithium7RF, zaccanti2009potassium39, lompe2010lithium6AD, lompe2010lithium6RF, nakajima2010lithium6, nakajima2011lithium6} (see also Ref.~\cite{mestrom2017jila} for a careful analysis of the experimental data). An additional challenge is to construct the simplest two-body models that are capable of reproducing the correct three-body physics. Separable potentials are often used for this purpose \cite{schmidt2012universality, naidon2014physicalorigin, naidon2014microscopic, Greene2016softcoreVdW, cong2016vdWaals, langmack2018finiteAbg}. A potential operator is separable when it can be written as
\begin{equation}
V_{sep} = \lambda \lvert g \rangle \langle g \rvert,
\end{equation}
where $\lambda$ represents the interaction strength. Atomic interactions are local, finite-range potentials, and therefore not separable. Nevertheless, separable potentials are extremely useful for studying three-body physics because the three-body equations for such an interaction are much easier to solve than those for a local finite-range potential.
The simplest separable potential is the contact interaction. This zero-range model has proven to be successful in predicting the universal scaling laws in Efimov physics that apply at large scattering lengths \cite{braaten2006universality}. In order to fix the three-body parameter of the Efimov spectrum that is still free in the zero-range model, separable models that give rise to finite-range effects can be used \cite{schmidt2012universality, naidon2014physicalorigin, naidon2014microscopic, Greene2016softcoreVdW, langmack2018finiteAbg}.
These separable finite-range interactions are also relevant for constructing many-body theories of quantum gases \cite{kohler2003atom_molecule_BEC} due to ease of implementation.  Recently, they were used to study the resonantly interacting Bose gas \cite{colussi2018dynamical}.
These developments raise the question of to what extent separable approximations for local potentials can be used for predicting three- and many-body properties. In this work, we investigate when the separable approximation breaks down on the three-body level and what is needed to improve on the separable approximation.

The necessity to go beyond theories based on separable interactions has been shown by a recent numerical study on the potential resonances of the Lennard-Jones potential \cite{mestrom2017jila}, revealing that the first excited Efimov trimer does not intersect the atom-dimer threshold. The authors of Ref.~\cite{mestrom2017jila} attributed this noncrossing of the first excited Efimov resonance to strong $d$-wave interactions near $a = 1$~$r_{vdW}$ \cite{dincao2012dwave}. Using the separable approximation for the van der Waals potential, Refs.~\cite{Greene2016softcoreVdW,cong2016vdWaals} considered only $s$-wave interactions and found that the first excited Efimov state does intersect the atom-dimer threshold. This contradiction suggests that a single-term separable approximation involving only $s$-wave interactions is insufficient in this particular case.


Here we study the effects of the nonseparability of local finite-range potentials on the Efimov spectrum using the two-body square-well interaction given by
\begin{equation} \label{sq:pot}
V_{SW}(r)=\begin{cases}-V_0&\mbox{$0\leq r<R$}\\0&\mbox{$r\geq R$},\end{cases}.
\end{equation}
where $r$ is the relative distance between the two particles and $R$ is the range of the potential. The advantage of the square-well potential lies in the fact that it is one of the simplest extensions of zero-range interaction models, incorporating finite-range effects in a well-defined way.  Additionally, many relevant two-body properties, such as the off-shell two-body $T$ matrix, are known analytically. This simplifies three-body calculations beyond the separable approximation of the off-shell two-body $T$ matrix when computing the Efimov states from the momentum-space representation of the Faddeev equations \cite{faddeev1993scattering}.
Another reason to study the square-well potential is that it does not belong to any of the classes of potentials that have been studied extensively in the context of Efimov physics \cite{naidon2014microscopic}.

We restrict ourselves to the case of identical spinless bosons and solve the three-body Faddeev equations for different potential depths $V_0$, which translates to a scattering length at a certain number of bound states in the two-body system. Efimov physics occurs near every potential resonance of the square-well interaction. For interactions strengths $V_0$ near these potential resonances, we solve the three-body Faddeev equations in the momentum-space representation by expanding the three-body wave function in spherical harmonics and functions that are related to the two-body bound and scattering states of the two-body potential. This expansion is directly related to a separable expansion of the potential itself and the corresponding two-body $T$ matrix. We discuss several existing expansion methods and analyze their advantages and disadvantages in calculations of the energies of the Efimov states. Our approach has the additional advantage that we can explicitly exclude or include $d$-wave interactions to study their effects, which is not easily possible in the position-based hyperspherical framework.  
Such effects have also not been investigated by momentum-space based studies that only involve $s$-wave separable interactions as a model for local finite-range potentials \cite{naidon2014microscopic,Greene2016softcoreVdW,cong2016vdWaals}.
We go beyond the separable approximation by fully expanding the square-well potential in separable terms, and we analyze the validity of the separable approach by comparing our results corresponding to the square-well potential with those corresponding to its separable approximation.




This paper is organized as follows. In \refSection{sec:Toff} we introduce the off-shell two-body $T$ matrix. In \refSection{sec:Faddeev} we review the Faddeev equations corresponding to bound states, three-body recombination and atom-dimer scattering. In \refSection{sec:SepExp} we outline and compare three methods for expanding the off-shell two-body $T$ matrix in terms that are separable in the incoming and outgoing momenta. These methods are used to calculate the properties of the Efimov spectrum for the square-well potential in \refSection{sec:Results}. Finally, we present the conclusions of our work in \refSection{sec:Conclusion}.

\section{The off-shell two-body $T$ matrix}\label{sec:Toff}
The Faddeev equations involve the two-body transition operator $T$ that we introduce in this section. It satisfies the Lippmann-Schwinger equation
\begin{equation} \label{eq:LippmannSchwinger}
T(z) = V + V G_0(z) T(z)
\end{equation}
where $G_0(z) = (z-H_0)^{-1}$ is the free Green's function, $V$ is the two-body interaction potential, $H_0$ is the two-body kinetic energy operator in the center-of-mass frame and $z$ is the complex energy of the two-particle system. The two-body $T$ matrix is then defined as $\langle \mathbf{p'} | T(z) | \mathbf{p} \rangle$ where $\mathbf{p}$ and $\mathbf{p}'$ are relative momenta of the two-particle system. Throughout this paper we use plane wave states that are normalized according to $\langle \mathbf{p'} | \mathbf{p} \rangle = \delta(\mathbf{p'}-\mathbf{p})$.

The two-body $T$ matrix is in general evaluated off the energy shell, which means that $p'^2 \neq p^2 \neq 2 \mu z$. Here the reduced mass of the two-body system is indicated by $\mu$ which equals $m/2$ for two identical particles of mass $m$. Since the energy and momentum of each two-particle subsystem is not conserved in a three-body system, the off-shell two-body $T$ matrix, which we will call the off-shell $T$ matrix for short, must be calculated in order to compute the energies of the Efimov states by using the Faddeev equations.

As we will see in \refSection{sec:Faddeev} we can reduce the dimensionality of the three-body integral equations by expanding the off-shell $T$ matrix. For spherically symmetric interactions, the $T$ matrix can be expanded in terms of Legendre polynomials $P_l$ as
\begin{equation} \label{T2matrixPartialWaveExp}
\langle \mathbf{p'} | T(z) | \mathbf{p} \rangle = \sum_{l = 0}^{\infty} (2 l + 1) P_l(\mathbf{\hat{p}'}\cdot \mathbf{\hat{p}}) t_l(p,p',z).
\end{equation}
We can further expand the off-shell partial-wave components $t_l(p,p',z)$ as a sum of terms that are separable in $p$ and $p'$, namely
\begin{equation} \label{tlexpansionGeneral}
t_l(p,p',z) = - \sum_{n = 1}^{\infty} \tau_{n l}(z) g_{n l}(p,z) g_{n l}(p',z),
\end{equation}
assuming real energies $z<0$.
There exist many ways in which this separable expansion can be done and whether the form factors $g_{n l}(p,z)$ are energy-dependent or not depends on the particular method used. In \refSection{sec:SepExp}, we will come back to the details of the separable expansion.

\refAppendixSection{app:Toff_SqW} presents the off-shell $T$ matrix for the square-well potential and relates it to the inherent potential resonances. There we also introduce the notation used in this paper for variables made dimensionless by scaling with the constants $R$, $m$, and $\hbar$; e.g., lengths, momenta and energies are expressed in finite-range units as $\bar{a} = a /R$, $\bar{p} = p R/\hbar$, and $\bar{E} = E m R^2/\hbar^2$, respectively.

\section{The Faddeev equations} \label{sec:Faddeev}

Here we review the Faddeev equations associated with bound states, three-body recombination, and atom-dimer scattering. We use these equations to find the Efimov spectrum and the corresponding three-body parameters in \refSection{sec:Results}. We solve the three-body equations in the momentum-space representation for three identical zero-spin bosons. Our three-body potential is the sum of three two-body interaction potentials, each of which is the square-well potential defined in \refEquation{sq:pot}.

\refSection{sec:FaddeevBS} presents the eigenvalue equation from which the bound trimer states are calculated. By expanding the three-body wave function in spherical harmonics and form factors, we obtain an integral equation that is solved numerically. When deeper two-body bound states exist, the eigenvalue equation will only have solutions for complex three-body energies $E$. However, without making the three-body energy complex, we can still estimate the energies of the three-body quasibound states as discussed in \refSection{sec:FaddeevResonances}. At the thresholds, i.e., $E= 0$ and $E = E_{2b}$, we can also extract information about the trimer states from three-body scattering properties such as the three-body recombination rate and the atom-dimer scattering length. Three-body recombination is an inelastic scattering event in which three free particles collide and a two-body bound state is formed. The free particle carries away part of the total momentum as the total momentum of the three-particle system is conserved. We present the equations to calculate the corresponding decay rates from the three-body transition operators for recombination in \refSection{sec:K3}. The three-body transition operators for atom-dimer scattering are presented in \refSection{sec:AtomDimer}. The atom-dimer scattering length can be calculated from the elastic atom-dimer transition operator and also gives the inelastic decay rate when more strongly bound dimer states exist.

\subsection{Three-body bound states}\label{sec:FaddeevBS}

Following Faddeev \cite{faddeev1993scattering}, the three-body bound states $\lvert \Psi \rangle$ can be calculated from
\begin{equation}
\lvert \Psi \rangle = - \sum_{\alpha = 1}^3 \frac{1}{E-H_0} \lvert \Phi_{\alpha} \rangle.
\end{equation}
where the vectors $\lvert \Phi_{\alpha} \rangle$ are determined by the following set of coupled equations:
\begin{equation} \label{OperatorPhi}
\lvert \Phi_{\alpha} \rangle = T_{\alpha}(E) G_0(E) \left(\lvert \Phi_{\beta} \rangle + \lvert \Phi_{\gamma} \rangle \right) \text{, } \alpha \beta \gamma = 1 2 3, 2 3 1, 3 1 2.
\end{equation}
Here $T_{\alpha}(E)$ is the two-body $T$ operator for scattering between particles $\beta$ and $\gamma$ in the presence of particle $\alpha$, \textit{i.e.\ } $T_{\alpha}(E) = V_{\alpha} + V_{\alpha} G_0(E) T_{\alpha}(E)$ where $G_0(E) = (E-H_0)^{-1}$ now contains the kinetic energy operators for all three particles in the center-of-mass frame. Now we define $\mathbf{q}_{\alpha}$ as the relative momentum of particle $\alpha$ with respect to the center of mass of the two-particle system $(\beta \gamma)$ and $\mathbf{p}_{\alpha}$ as the relative momentum between particles $\beta$ and $\gamma$. In momentum space, the operators $T$ and $T_{\alpha}$ are then related by
\begin{equation}
\langle \mathbf{p}_{\alpha}, \mathbf{q}_{\alpha} | T_{\alpha}(E) | \mathbf{p}_{\alpha}', \mathbf{q}_{\alpha}' \rangle = \langle \mathbf{q}_{\alpha} | \mathbf{q}_{\alpha}' \rangle \langle \mathbf{p}_{\alpha} | T(E-\frac{3}{4 m}q_{\alpha}^2) | \mathbf{p}_{\alpha}' \rangle.
\end{equation}

For three identical spinless bosons the set of equations given in \refEquation{OperatorPhi} reduces to a single integral equation. In the momentum-space representation, we obtain the following three-body equation
\begin{equation}\label{3bodyEq}
\begin{split}
\langle \textbf{p}, \textbf{q} | \Phi(E) \rangle &= 
 \int d\textbf{q}' \frac{t \left(\textbf{p}, \frac{1}{2} \textbf{q} + \textbf{q}', E - \frac{3}{4 m} q^2  \right)}{E-\frac{1}{m}\left(q^2+ \textbf{q}\cdot \textbf{q}' + q'^2 \right)} \\
 &\langle \textbf{q} + \frac{1}{2}\textbf{q}', \textbf{q}' | \Phi(E) \rangle,
 \end{split}
\end{equation}
where we have dropped the index $\alpha$ and have defined the symmetrized two-body $T$ matrix, $t \left(\mathbf{p}, \mathbf{p}', z \right)$, as \cite{schadow2000STM}
\begin{equation}
t \left(\mathbf{p}, \mathbf{p}', z \right) = \langle \mathbf{p} | T(z) | \mathbf{p}' \rangle + \langle \mathbf{p} | T(z) | -\mathbf{p}' \rangle,
\end{equation}
which only includes partial-wave components with even values of the angular momentum quantum number $l$.  The next step is to apply a partial-wave expansion \cite{kharchenko1969separable,sitenko1991scattering} to \refEquation{3bodyEq} for total angular momentum $\mathbf{L = 0}$, which is allowed by conservation of total angular momentum, i.e.,
\begin{equation}
\langle \textbf{p}, \textbf{q} | \Phi(E) \rangle = \sum_{l=0}^{\infty} (2 l + 1) P_l(\hat{\mathbf{p}}\cdot \hat{\mathbf{q}}) \tilde{\Phi}_{l}(p,q,E),
\end{equation}
so that \refEquation{3bodyEq} reduces to
\begin{equation}\label{3bodyEqPartialWaveExp}
\begin{aligned}
\tilde{\Phi}_l&(p,q,E)  = \int d\mathbf{q}' \frac{1}{E-\frac{1}{m}\left(q^2+ \mathbf{q}\cdot \mathbf{q}' + q'^2 \right)} \\
&\left(2 \Delta_{l} P_l(\hat{\mathbf{q}} \cdot \reallywidehat{\frac{1}{2} \mathbf{q} + \mathbf{q}'})
t_l(p,|\frac{1}{2} \mathbf{q} + \mathbf{q}'|,E - \frac{3}{4 m} q^2)\right)\\ 
&\sum\limits_{l'=0}^\infty (2 l' + 1) P_{l'}(\reallywidehat{\mathbf{q} + \frac{1}{2}\mathbf{q}'} \cdot \reallywidehat{\mathbf{q}'})\tilde{\Phi}_{l'}(|\mathbf{q} + \frac{1}{2}\mathbf{q}'|,q',E) 
\end{aligned}
\end{equation}
where $\Delta_{l} = \frac{1}{2}\left(1+(-1)^l\right)$. Therefore all components $\tilde{\Phi}_l(p,q,E)$ with odd $l$ are equal to zero. This set of equations is an infinite set of two-dimensional integral equations. Most papers in which the energies of the Efimov trimers are calculated by using the Faddeev equations \cite{cong2016vdWaals, Greene2016softcoreVdW, naidon2014physicalorigin, naidon2014microscopic} use a separable approximation of the $s$-wave partial-wave component $t_0(p,p',E)$ and neglect the interactions involving $l = 2, 4, 6, ...$, so that \refEquation{3bodyEqPartialWaveExp} reduces to a single one-dimensional integral equation. However, we transform \refEquation{3bodyEqPartialWaveExp} to an infinite set of one-dimensional integral equations by substituting the separable expansion given by \refEquation{tlexpansionGeneral} into \refEquation{3bodyEqPartialWaveExp}. For this purpose, we also define the quantities $\tilde{\phi}_{l n}(q,E)$ such that the functions $\tau_{n l}\left(E-3q^2/(4m)\right) \tilde{\phi}_{l n}(q,E)$ are the expansion coefficients of $\tilde{\Phi}_l(p,q,E)$ with respect to the basis $\left\{g_{n l}\left(p,E-3 q^2/(4 m)\right)\right\}$, i.e., 
\begin{equation}
\tilde{\Phi}_l(p,q,E) = \sum_{n = 1}^{\infty} g_{n l}\left(p,Z_q\right) \tau_{n l}\left(Z_q\right) \tilde{\phi}_{l n}(q,E),
\end{equation}
where $Z_{q} = E - 3 q^2/(4 m)$.
With those definitions, it can be derived that the resulting three-body equation is
\begin{equation} \label{3bodyCoupledEq}
\begin{aligned}
\tilde{\phi}_{l n}(q,E) &= - \int d\mathbf{q}' \frac{2 \Delta_{l} P_l(\hat{\mathbf{q}} \cdot \reallywidehat{\frac{1}{2} \mathbf{q} + \mathbf{q}'})}{E-\frac{1}{m}\left(q^2+ \mathbf{q}\cdot \mathbf{q}' + q'^2 \right)} \\
&\tau_{n l}(Z_{q'}) g_{n l}(|\frac{1}{2} \mathbf{q} + \mathbf{q}'|,Z_{q})\\
&\sum\limits_{l'=0}^\infty \sum_{n' = 1}^{\infty} (2 l' + 1) \Delta_{l'} P_{l'}(\reallywidehat{\mathbf{q} + \frac{1}{2}\mathbf{q}'} \cdot \reallywidehat{\mathbf{q}'}) \\
&g_{n' l'}(|\mathbf{q} + \frac{1}{2}\mathbf{q}'|,Z_{q'}) \tilde{\phi}_{l' n'}(q',E).
\end{aligned}
\end{equation}
Here we have also assumed that an orthonormality condition for the form factors $g_{n l}(p,z)$ exists, which is the case for the separable expansions considered below. This infinite set of coupled one-dimensional integral equations reduces to a finite set of equations when a finite number of terms is used to expand the off-shell components $t_l(p,p',z)$. \refEquation{3bodyCoupledEq} is solved by discretizing the momenta $q$ and $q'$, so that this coupled set of equations can be written as one matrix equation. The matrix representation of the collection of integral operators in \refEquation{3bodyCoupledEq} is called the kernel in the following. The kernel has an eigenvalue equal to 1 at energies where a bound trimer state exists. Solutions can be found by varying the three-body energy $E$ and the scattering length $a$, where the latter is varied through changing the depth of the square well.

\subsection{Three-body resonances}\label{sec:FaddeevResonances}

The previous section has dealt with three-body bound states consisting of three identical spinless bosons that only exist below the two-body ground-state energy $E_{2b,0}$. For $E_{2b,0}<E<0$ the solutions to \refEquation{OperatorPhi} correspond to three-body resonances. Therefore solutions only exist for complex energies, $E = E_R - i/2 \ \Gamma$ where $E_R$ and $\Gamma$ are real.

Numerically, we do not search in the complex energy plane to find an eigenvalue that equals one, but insert real energies into \refEquation{3bodyCoupledEq} and search for eigenvalues whose real part equals one. This method is expected to work well if the complex part of the eigenvalue is small compared to the real part, so that the real part $E_R$ of the Efimov resonance is not much affected by the complex part of the eigenvalue of the kernel. Since the functions $\tau_{n l}(Z_{q'})$ contain singularities on the integration contour when considering energies $E>E_{2b,0}$, we have to deform our integration contour near the singularities. This method is equivalent to plugging in complex energies $E \pm i \epsilon$ where $\epsilon \rightarrow 0$. The deformation of the contour can be performed most easily by splitting the integral into a principal value integral along the real axis and a complex part that is proportional to the residue of the integrand.

The validity of this approach to calculate the Efimov resonances is tested by comparing the corresponding results at $E = 0$ with the results obtained for three-body recombination of which the formalism is discussed in \refSection{sec:K3}. An advantage of the eigenvalue approach over the recombination rate analysis discussed below is that such a simple method would also be applicable for negative energies.

\subsection{Three-body recombination}
\label{sec:K3}

The value of the three-body parameter $a_{-}$ can also be calculated from the maxima of the low-energy three-body recombination rate $K_3$ \cite{esry1999recombination}, which rapidly grows near resonance with increasing scattering length as $a^4$ \cite{fedichev1996K3shallow, nielsen1999recombination, esry1999recombination}. The rate of decrease in the number density $n$ of a thermal cloud of atoms due to three-body recombination is given by
\begin{equation}
\frac{d n}{dt} = - \frac{1}{2} K_3 n^3,
\end{equation}
where the definition of $K_3$ is consistent with Ref.~\cite{braaten2008recombination}.

Several approaches can be adopted to calculate the three-body recombination rate. One method is to use the adiabatic hyperspherical approach \cite{esry1999recombination, suno2002ThreeBodyRecomb} which solves the three-body Schr\"{o}dinger equation in position space to obtain the $S$-matrix elements for three-body recombination. In this paper we use the Alt-Grassberger-Sandhas (AGS) equations \cite{alt1967ags},
\begin{equation} \label{eq:AGS_recombination}
U_{\alpha 0} = G_0^{-1} + \sum_{\substack{\beta = 1 \\ \beta \neq \alpha}}^{3} T_{\beta} G_0 U_{\beta 0},
\end{equation}
to find the transition amplitude for three-body recombination. Here $\alpha = 1, 2, 3$ labels the three possible configurations for the atom-dimer state, and all operators depend on the three-body energy for which we take the complex energy $E + i 0$, indicating that the real three-body energy $E$ is approached from the upper half of the complex energy plane. In the following, we also assume the energy dependence of the operators to be implicit for notational compactness.

For identical particles, it is convenient to define the operator $\breve{U}_{\alpha 0} \equiv T_{\alpha} G_0 U_{\alpha 0} (1+P)$, where $P$ is the sum of the cyclic and anticyclic permutation operators. We derive from \refEquation{eq:AGS_recombination} that the operator $\breve{U}_{\alpha 0}$ satisfies the inhomogeneous equation
\begin{equation}\label{eq:Ubreve}
\breve{U}_{\alpha 0} = T_{\alpha} (1 + P) + T_{\alpha} G_0 P \breve{U}_{\alpha 0},
\end{equation}
which we solve by direct matrix inversion in the momentum-space representation. If there is one eigenvalue of the operator $T_{\alpha} G_0 P$ close to one, the solution of \refEquation{eq:Ubreve} is dominated by the corresponding eigenvector and the transition amplitude is large. This shows that the search for eigenvalues close to one should give good approximations where the maxima in the three-body recombination rate can be found, as suggested in \refSection{sec:FaddeevResonances}.

In order to calculate the three-body recombination rate, we need to relate the operators $\breve{U}_{\alpha 0}$ to the on-shell transition amplitudes.
From \refEquation{eq:AGS_recombination} we derive that
\begin{equation}
U_{\alpha 0}(1+P) = G_0^{-1} (1 + P) + P \breve{U}_{\alpha 0},
\end{equation}
so that the zero-energy on-shell transition amplitude can be calculated from
\begin{equation}\label{eq:Ua0_zero_E_limit}
\begin{aligned}
\lim_{E \rightarrow 0} {}_{\alpha}\langle \mathbf{q}_d, \varphi_d | &U_{\alpha 0}(E) | \mathbf{q}_0, \mathbf{p}_0 \rangle \\
&= \frac{1}{3} \lim_{E \rightarrow 0} {}_{\alpha}\langle \mathbf{q}_d, \varphi_d | P\breve{U}_{\alpha 0}(E) | \mathbf{q}_0, \mathbf{p}_0 \rangle, 
\end{aligned}
\end{equation}
The state $\lvert \mathbf{q}_{d}, \varphi_d \rangle_{\alpha}$ consists of a two-body bound state $\lvert \varphi_d \rangle$ formed by particles $\beta$ and $\gamma$ and a free particle $\alpha$ whose relative momentum is $\mathbf{q}_{d}$ with respect to the center of mass of the dimer.
The requirement that the transition amplitude ${}_{\alpha}\langle \mathbf{q}_d, \varphi_d | U_{\alpha 0}(E) | \mathbf{q}_0, \mathbf{p}_0 \rangle$ is evaluated on the energy shell implies that $E =~E_{2b,d}~+~3/(4 m)~q_d^2 =~p_0^2/m~+~3/(4 m)~q_0^2$, where  $E_{2b,d}$ is the bound-state energy of the dimer state labeled by the quantum number $d$.

Since we are only interested in the values of the three-body parameter $a_{-}$, it is sufficient to calculate the recombination rate at zero energy, which leads to a couple of simplifications in the numerical implementation. At positive energies the calculation of the transition amplitudes ${}_{\alpha}\langle \mathbf{q}_d, \varphi_d | U_{\alpha 0}(E) | \mathbf{q}_0, \mathbf{p}_0 \rangle$ is hard due to the singularity resulting from the free Green's operator $G_0$. For zero energy there is no singularity, which makes the calculation of the transition amplitudes much simpler. Furthermore, in the zero-energy limit three identical particles can only recombine for zero total angular momentum \cite{esry2001thresholdlaws}, which also simplifies the calculation.

The three-body recombination rate at zero energy is determined from the on-shell transition amplitudes ${}_{\alpha}\langle \mathbf{q}_d, \varphi_d | U_{\alpha 0}(E) | \mathbf{q}_0, \mathbf{p}_0 \rangle$ by the following formula \cite{smirne2007K3rubidiumBEC, lee2007excited}:
\begin{equation}\label{eq:K3zerovsU}
\begin{aligned}
K_3(0) &= \frac{24 \pi m}{\hbar} (2 \pi \hbar)^6 \sum_{d = 1}^{N_d} \int \,d\hat{\mathbf{q}}_d \\
& \lim_{E \rightarrow 0} q_d \cdot \lvert {}_{\alpha}\langle \mathbf{q}_d, \varphi_d | U_{\alpha 0}(E) | \mathbf{q}_0, \mathbf{p}_0 \rangle \rvert^2,
\end{aligned}
\end{equation}
where $N_d$ is the number of dimer states $\lvert \varphi_d \rangle$ supported by the potential. Energy conservation determines the final relative momentum $q_d$ by $|E_{2b,d}| = \frac{3}{4 m} q_d^2$. \refAppendixSection{app:AGS_integral_eq} provides more details about the way we calculate $K_3(0)$ exploiting the separable expansion of the two-body $T$ matrix.

The zero-energy three-body recombination rate behaves universally at scattering lengths $a$ that are much larger than the range of the two-body interaction potential \cite{braaten2000K3, braaten2001K3deep, braaten2004atomdimer, braaten2006universality, braaten2008recombination}. For large negative scattering lengths the behavior of $K_3$ is given by
\begin{equation}\label{eq:K3universalNega0}
K_3(0) = 6 C_{-} \frac{\sinh(2 \eta_{*})}{\sin^2\left(s_0 \ln(a/a_{-})\right) + \sinh^2(\eta_{*})} \frac{\hbar a^4}{m} 
\end{equation}
where $C_{-} \approx 4590$ and $s_0 \approx 1.00624$ \cite{braaten2004atomdimer, braaten2006universality}. The inelasticity parameter $\eta_{*}$ determines the probability to decay to deeply bound molecules according to $(1 - e^{- 4 \eta_{*}})$ \cite{dincao2018review}. This parameter thus controls the width of the Efimov resonances. Finite-range corrections to the universal expressions for $K_3(0)$ have been investigated in Ref.~\cite{ji2012rangecorrections, ji2015rangecorrections, dincao2018review}. After calculating $K_3(0)$ from \refEquation{eq:K3zerovsU} we fit the data near the three-body resonance with \refEquation{eq:K3universalNega0} to obtain the value of $a_{-}$ and the loss parameter $\eta_{*}$.

\subsection{Atom-dimer scattering}\label{sec:AtomDimer}

The $(n+1)$th Efimov trimer merges with the atom-dimer threshold at a scattering length $a = a_{*,n}$ where $n = 0, 1, 2, ...$. These values can be determined from the maxima of the low-energy inelastic atom-dimer scattering rate $\beta$ \cite{braaten2004atomdimer,braaten2006universality}, which decreases the atom density $n_A$ and dimer density $n_D$ in a trap according to
\begin{equation}
\frac{d n_D}{dt} = \frac{d n_A}{dt} = - \beta n_D n_A.
\end{equation}
The loss rate coefficient $\beta$ is related to the imaginary part of the elastic atom-dimer scattering length $a_{ad}$ \cite{braaten2004atomdimer,braaten2006universality} by
\begin{equation}
\beta = -\frac{6 \pi \hbar}{m} \text{Im}(a_{ad}).
\end{equation}

To determine the three-body parameter $a_{*}$ at positive scattering lengths we therefore calculate the elastic atom-dimer scattering amplitude at zero energy. This amplitude diverges whenever a trimer state merges with the atom-dimer threshold. Again our starting point is the AGS approach \cite{alt1967ags}. Similarly to \refEquation{eq:AGS_recombination}, the transition operators for atom-dimer rearrangement are determined by the following system of equations:
\begin{equation} \label{eq:AGS_rearrangement}
U_{\gamma \alpha} = (1 - \delta_{\gamma \alpha})G_0^{-1} + \sum_{\substack{\beta = 1 \\ \beta \neq \gamma}}^{3} T_{\beta} G_0 U_{\beta \alpha},
\end{equation}
where we assumed the dependence on the three-body energy $E$ to be implicit, for which the limit to real values is taken from the upper half of the complex energy plane.
The total transition amplitude for atom-dimer scattering from the incoming atom-dimer state $ \lvert \mathbf{q}_i, \varphi_i \rangle$ to the outgoing atom-dimer state $\lvert \mathbf{q}_f, \varphi_f \rangle_{\gamma}$ is given by
\begin{equation}\label{eq:AD_transition_amplitude}
\sum_{\alpha = 1}^{3}  {}_{\gamma}\langle \mathbf{q}_f, \varphi_f | U_{\gamma \alpha}(E) | \mathbf{q}_i, \varphi_i \rangle_{\alpha} \equiv {}_{\gamma}\langle \mathbf{q}_f, \varphi_f | U_{\gamma}^{i}(\mathbf{q}_i, E) \rangle,
\end{equation}
which we use to define the states $\lvert U_{\gamma}^{i}(\mathbf{q}_i, E) \rangle$. The energies are evaluated on the energy shell, which means that $E = E_{2b,i}~+~3/(4 m)~q_i^2 = E_{2b,f}~+~3/(4 m)~q_f^2$. The summation over the initial atom-dimer configurations in \refEquation{eq:AD_transition_amplitude} is needed to properly account for the identical nature of the particles.  From general scattering theory \cite{taylor1972scattering}, it can be derived that the $s$-wave atom-dimer scattering length $a_{ad}$ is related to the on-shell transition amplitude for elastic atom-dimer scattering by
\begin{equation} \label{eq:aAD_related_to_U}
a_{ad} = \frac{8}{3} \pi^2 m \hbar \lim_{E \to E_{2b,i}}
\sum_{\alpha = 1}^{3} 
 {}_{\gamma}\langle \mathbf{q}_f, \varphi_i | U_{\gamma \alpha}(E) | \mathbf{q}_i, \varphi_i \rangle_{\alpha},
\end{equation}
where $i$ labels the considered dimer state. The momenta $q_i$ and $q_f$ also go to zero in the limit $E\to 0$ since we are considering the on-shell transition amplitude. Note that even though the magnitudes of the momenta $\mathbf{q}_i$ and $\mathbf{q}_f$ are the same on the energy shell, the orientations of these vectors do not need to be the same.

From \refEquation{eq:AGS_rearrangement} we derive that the states $\lvert U_{\gamma}^{i}(\mathbf{q}_i, E) \rangle$, defined by \refEquation{eq:AD_transition_amplitude}, are determined from
\begin{equation} \label{eq:AD_Ugamma}
\lvert U_{\gamma}^{i}(\mathbf{q}_i, E) \rangle = P G_0^{-1} \lvert \mathbf{q}_i, \varphi_i \rangle_{\gamma} + P \lvert \tilde{U}_{\gamma}^{i}(\mathbf{q}_i, E) \rangle,
\end{equation}
where we have defined some new states $\lvert \tilde{U}_{\gamma}^{i}(\mathbf{q}_i, E) \rangle \equiv T_{\gamma} G_0 \lvert U_{\gamma}^{i}(\mathbf{q}_i, E) \rangle$. The states $\lvert \tilde{U}_{\gamma}^{i}(\mathbf{q}_i, E) \rangle$ are therefore determined from the inhomogeneous equation
\begin{equation} \label{eq:AD_tildeUgamma}
\lvert \tilde{U}_{\gamma}^{i}(\mathbf{q}_i, E) \rangle = T_{\gamma} P \lvert \mathbf{q}_i, \varphi_i \rangle_{\gamma} + T_{\gamma} G_0 P \lvert \tilde{U}_{\gamma}^{i}(\mathbf{q}_i, E) \rangle,
\end{equation}
which we again solve by direct matrix inversion in the momentum-space representation. To reduce the dimensionality of this equation we expand it in terms of spherical harmonics and we also expand the $T$ operator in separable terms. The resulting equations to calculate the $s$-wave atom-dimer scattering length can be found in \refAppendixSection{app:AGS_integral_eq}.

Universal expressions also exist for the atom-dimer scattering length and the corresponding loss rate coefficient $\beta$ \cite{braaten2004atomdimer,braaten2006universality}. These are given by
\begin{align}
a_{ad} &= \left(1.46 + 2.15 \cot\left[s_0 \text{ln}(a/a_{*}) + i \eta_{*} \right] \right) a \text{ and} \label{eq:aAD_Universal}\\
\beta &= \frac{20.3 \sinh(2 \eta_{*})}{\sin^2\left(s_0 \ln(a/a_{*})\right) + \sinh^2(\eta_{*})} \frac{\hbar a}{m}, \label{eq:beta_Universal}
\end{align}
which are valid for large positive scattering lengths. So we calculate $a_{ad}$ from \refEquation{eq:AD_ScatL} after which the data are fitted with \refEquation{eq:aAD_Universal} to obtain the value of $a_{*}$ and the loss parameter $\eta_{*}$.

\section{Separable expansions of the off-shell $T$ matrix}\label{sec:SepExp}

In the previous section we discussed the three-body equations that we solve to find the Efimov spectrum and the corresponding three-body parameters. We simplified the equations by approximating the two-body interaction in separable terms. Here we discuss several approaches that can be used to expand the partial-wave components of the off-shell $T$ matrix in a series of terms that are separable in the initial and final momenta. First, we analyze two expansion methods that can be used to expand the partial-wave components of the off-shell $T$ matrix in separable terms without having cross terms as in \refEquation{tlexpansionGeneral}. Both expansions, \textit{i.e.\ } the spectral representation and the Weinberg expansion, converge to the right two-body $T$ matrix. These two methods have properties that are similar with respect to the two-body bound states. The difference in computation time in three-body calculations using either method depends on the details of these calculations. Additionally, we discuss the EST expansion method resulting in a separable expansion including cross terms. Nevertheless, this method is capable of producing a single-term separable approximation for the two-body $T$ matrix that can give reasonable results for the three-body parameter \cite{naidon2014physicalorigin, naidon2014microscopic}. 

Throughout this section, we only present separable expansions of $t_l(p,p',z)$ for real momenta $p$ and $p'$ and real energies $z \leq 0$. Since we only solve the Faddeev equations for three-body energies $E\leq 0$, the energy variable $z$ in $t_l(p,p',z)$ only takes on values smaller than or equal to zero. Singularities of the kernel of the three-body integral equations that are located on the real energy axis are handled via the residue theorem as discussed in \refSection{sec:FaddeevResonances}, so that we do not need to deform our integration contour.
Finally, we note that the methods considered in this section are valid when regular scattering theory is valid, which implies that the potential is of short-range nature; i.e., it should fall off sufficiently fast with increasing interparticle separation \cite{taylor1972scattering}.

\subsection{Method I: the spectral representation}
The first method that we describe can be used when the off-shell $T$ matrix is known. 
Since we only consider real energies $z \leq 0$, the partial-wave components $t_l(p,p',z)$ are real.  
The form factors $g_{n l}(p,z)$ can be defined as the solutions of the following integral equation:
\begin{equation} \label{M1_DefFormFactors}
- \int_{0}^{\infty} t_l(p,p',z) g_{n l}(p',z) dp' = \tau_{n l}(z) g_{n l}(p,z).
\end{equation}
The index $n$ labels the eigenvalues and corresponding eigenvectors. Since the kernel $t_l(p,p',z)$ is symmetric in $p$ and $p'$, the eigenvalues $\tau_{n l}(z)$ are real \cite{pipkin1991integralequations}. Furthermore, the eigenvectors $g_{n l}(p,z)$ corresponding to different eigenvalues are orthogonal and eigenvectors corresponding to the same eigenvalue can be orthogonalized \cite{pipkin1991integralequations}. The orthonormalization condition is given by
\begin{equation} \label{M1_orthornormal}
\int_{0}^{\infty} g_{n' l}(p,z) g_{n l}(p,z) dp = \delta_{n' n}
\end{equation}
for real energies $z$.
The spectral representation of $t_l(p,p',z)$ is then given by \refEquation{tlexpansionGeneral}.

\subsection{Method II: the Weinberg expansion}
Another method to obtain a separable expansion is based on the Hilbert-Schmidt theorem for symmetric integral equations \cite{kharchenko1969separable,sitenko1991scattering, schmid1974threebody}. This approach has been used first by Weinberg \cite{weinberg1963expansion} to eliminate the divergence of the Born series and is also known as the quasiparticle method or the Weinberg series. The starting point of this method is to define the vectors $\lvert g(z)\rangle$ as the eigenfunctions of the operator $V G_0(z)$ with eigenvalue $\eta(z)$, i.e.,
\begin{equation} \label{eq:DefgstatesVG0method}
V G_0(z) \lvert g(z)\rangle = \eta(z) \lvert g(z)\rangle.
\end{equation}
For $z<0$ the eigenfunctions of the operator $V G_0$ are related to the two-body bound state wave functions $\lvert \phi(z)\rangle$ of the energy-dependent potential $V/\eta(z)$ by
\begin{equation} \label{eq:MethodVG0DefPhi}
\lvert \phi(z)\rangle = N_{norm} G_0(z) \lvert g(z)\rangle
\end{equation}
where $N_{norm}$ is a normalization constant. Now \refEquation{eq:DefgstatesVG0method} can be rewritten in the momentum representation. By using the partial-wave expansion of the potential, namely
\begin{equation} \label{VPartialWaveExp}
\langle \mathbf{p'} | V | \mathbf{p} \rangle = \sum_{l = 0}^{\infty} (2 l + 1) P_l(\mathbf{\hat{p}'}\cdot \mathbf{\hat{p}}) V_l(p,p'),
\end{equation}
and writing the functions $\langle \mathbf{p} | g(z)\rangle$ as $\langle \mathbf{p} | g_{n l m}(z)\rangle =  Y_l^m(\mathbf{\hat{p}}) g_{n l}(p,z)$, we end up with
\begin{equation} \label{eq:DefFormFactorsVG0method}
- 4 \pi \int_{0}^{\infty} V_l(p,p') \frac{1}{\frac{p'^2}{2 \mu}-z}  g_{n l}(p',z) p'^2 \,dp' = \eta_{n l}(z) g_{n l}(p,z).
\end{equation}
We label the eigenvalues $\eta_{n l}(z)$ in decreasing order of their absolute values. The form factors $g_{n l}(p,z)$ are orthogonal through the orthonormalization condition given by
\begin{equation}  \label{eq:orthornormalVG0method}
\int_{0}^{\infty}g_{n' l}(p,z) g_{n l}(p,z) \frac{1}{\frac{p^2}{2 \mu}-z} p^2 dp = \delta_{n' n}.
\end{equation}
The eigenvalues $\eta_{n l}(z)$ and form factors $g_{n l}(p,z)$ are real for real, negative energies $z$ \cite{sitenko1991scattering}. Furthermore, the partial-wave components $t_l(p,p',z)$ can be represented in separable terms by \refEquation{tlexpansionGeneral} where the expansion coefficients are given by \cite{kharchenko1969separable,sitenko1991scattering}:
\begin{equation} \label{eq:tauVG0}
\tau_{n l}(z) = \frac{1}{4 \pi} \frac{ \eta_{n l}(z)}{1- \eta_{n l}(z)},
\end{equation}
while the energy-independent components of the potential can be expanded as
\begin{equation} \label{eq:VlExpansionVG0}
V_l(p,p') = - \frac{1}{4 \pi}  \sum_{n = 1}^{\infty} \eta_{n l}(z)g_{n l}(p,z) g_{n l}(p',z).
\end{equation}
Eqs.~(\ref{eq:tauVG0})~and~(\ref{eq:VlExpansionVG0}) can be derived from the two-body Lippmann-Schwinger equation, \refEquation{eq:LippmannSchwinger}, combined with the definition of the form factors, \refEquation{eq:DefFormFactorsVG0method}, and the corresponding orthonormality condition given by \refEquation{eq:orthornormalVG0method}. 

\begin{figure}[hbtp]
    \centering
    \includegraphics[width=3.4in]{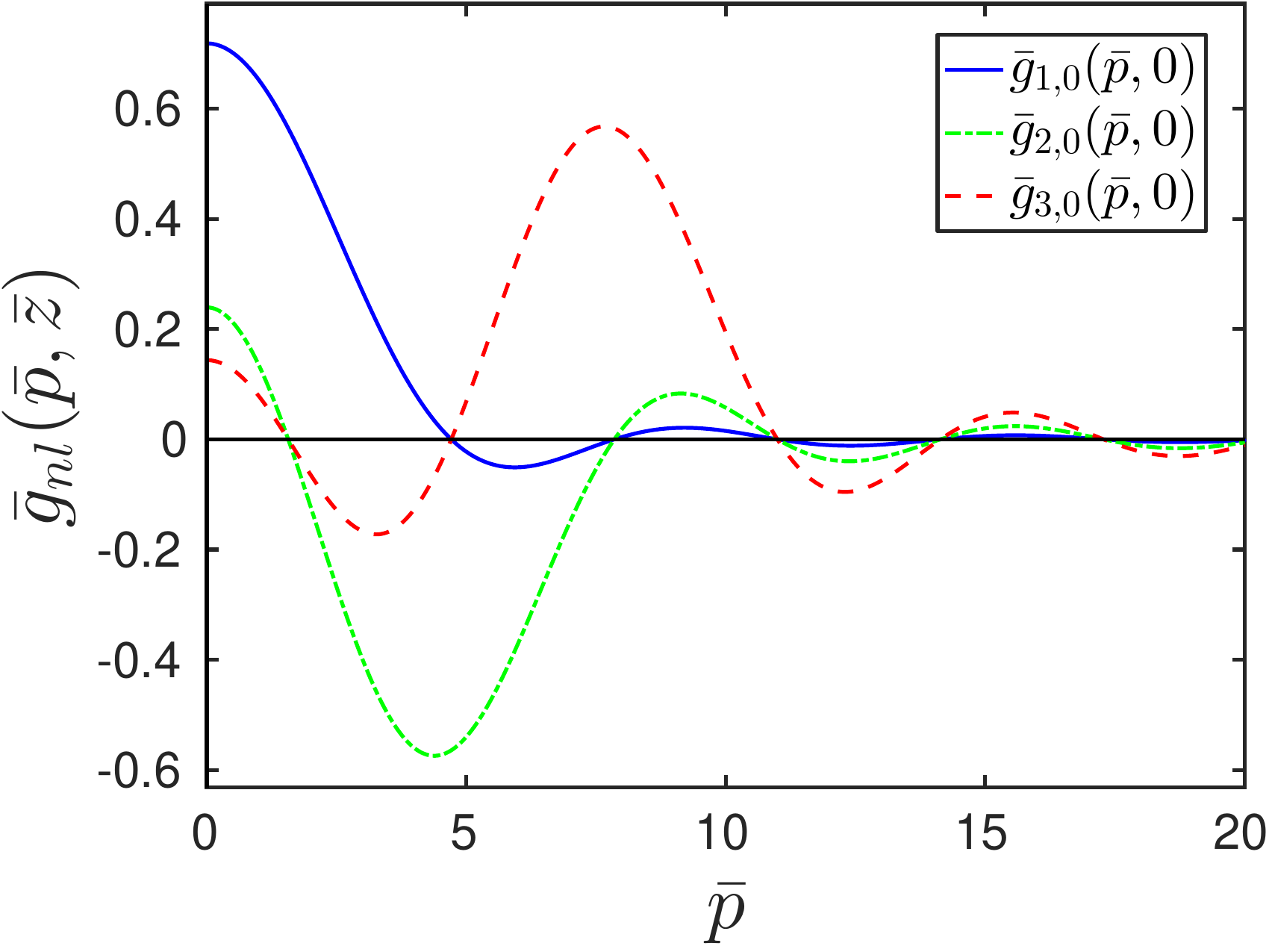}
    \caption{The dimensionless form factors $\bar{g}_{n l}(\bar{p},\bar{z}) \equiv g_{n l}(p,z) \sqrt{m \hbar/R}$ of the square-well potential calculated at $z = 0$ for $l = 0$ and $n = 1, 2, 3$ using the Weinberg expansion method.}
    \label{fig:FormFactors123_VG0_Comparison_q2_0}
\end{figure}

For the square-well potential, the eigenfunctions $g_{n l}(p,z)$ and eigenvalues $\eta_{n l}(z)$ can be found analytically for energies $z\leq 0$ \cite{sitenko1991scattering}. Fig.~\ref{fig:FormFactors123_VG0_Comparison_q2_0} shows these form factors for $n = 1, 2, 3$ and $l=0$ calculated at $z = 0$. Clearly, the form factors are oscillating functions which converge to zero as the magnitude of $p$ increases. Furthermore, for $z\leq 0$ all eigenvalues $\eta_{n l}(z)$ will be positive, so that Mercer's theorem \cite{hilbert1953mathematics} applies to the symmetrized kernel of \refEquation{eq:DefFormFactorsVG0method}. From this theorem it can be proven that the series in \refEquation{eq:VlExpansionVG0} converges absolutely and uniformly and so does the series of \refEquation{tlexpansionGeneral}. We have confirmed for the square-well potential that the Weinberg expansion indeed converges to the analytical $T$ matrix given by \refEquation{ToffmomentumSqW} for all negative energies relevant for the three-body calculations. However, for large negative energies the convergence of the Weinberg expansion is slow, because the eigenvalues $\eta_{n l}(z)$ decrease with increasing $|z|$ \cite{kharchenko1969separable}. The slow convergence for large negative $z$ poses no problem however since that energy regime is not of great relevance for the calculation of the weakly bound Efimov states.

Another method that we would like to mention is the unitary pole expansion (UPE), which was first suggested by Harms \cite{harms1970upe} and which is just a special case of the Weinberg expansion discussed above. The energy-dependent form factors defined by the Weinberg expansion can be made energy-independent if one fixes the energy $z$ in \refEquation{eq:DefgstatesVG0method} at some constant $E_b$. This means that the form factors are defined by
\begin{equation}
V G_0(E_b) \lvert g(E_b)\rangle = \eta(E_b) \lvert g(E_b)\rangle.
\end{equation}
For Efimov physics we are mainly interested in energies close to zero, so that it is natural to choose $E_b = 0$. The single-term approximation of the UPE is called the unitary pole approximation (UPA) \cite{lovelace1964upa,fuda1968upa}. In this paper we will not use the UPE because this expansion cannot be written in the form of \refEquation{tlexpansionGeneral} for $z\neq E_b$.

\subsection{Method III: the EST expansion} \label{ssec:EST}
The final method that we consider here is the EST method \cite{est1973}, which can also be used to approximate the partial-wave components $t_l(p,p',z)$ by a separable expansion. Since one disadvantage of this method is that it is difficult to obtain the next separable term in the expansion of $t_l(p,p',z)$ \cite{lehman1998separable}, most calculations involving the EST method take only the first separable term into account. This single-term approximation is a generalization of the unitary pole approximation. It provides a separable potential that reproduces one eigen- or scattering state of the original Hamiltonian at a specific energy $\tilde{E}$. For energies $\tilde{E}<0$, this separable approximation is identical to the UPA. For energies $\tilde{E}\geq 0$ the definition of the form factors $\lvert g \rangle$ is given by
\begin{equation}\label{EST:eq:Definition_gstates}
\lvert g \rangle = V \lvert \psi^{(+)}_{\tilde{E}} \rangle,
\end{equation} 
where $\lvert \psi^{(+)}_{\tilde{E}}\rangle$ is the considered scattering state. The separable potential $V_{sep} = \lambda \lvert g \rangle  \langle g \rvert$ will then reproduce the same scattering state $\lvert \psi^{(+)}_{\tilde{E}} \rangle$ as the nonseparable potential $V$ if the strength $\lambda$ is chosen to be
\begin{equation} \label{EST:eq:condition2_v3}
\lambda =  \left(\langle \psi^{(+)}_{\tilde{E}} |  V | \psi^{(+)}_{\tilde{E}} \rangle\right)^{-1}.
\end{equation}

The EST method has been used earlier to calculate the energies of Efimov states corresponding to the potential resonances of some van der Waals potentials \cite{naidon2014physicalorigin, naidon2014microscopic, cong2016vdWaals}. In those references the separable potential is chosen such that it reproduces the zero-energy $s$-wave scattering state of the considered two-body van der Waals potentials. Choosing $\tilde{E} = 0$ results in a scattering length of the separable potential that is the same as the one of the original potential $V$. This suggests that this method is especially useful for calculations of the three-body parameter $a_{-}$. This is an important advantage of the EST approach over other separable approximations such as the single-term approximation of methods I and II.

Once we have calculated the form factors from \refEquation{EST:eq:Definition_gstates}, we can find the separable approximation to the two-body $T$ matrix. The separable approximation $V_{sep}(p,p') = \lambda g(p) g(p')$ for the partial-wave component $V_0(p,p')$ and the ansatz $t_0(p,p',z) = \tau(z) g(p) g(p')$ can be substituted in the Lippmann-Schwinger equation for $t_0(p,p',z)$, which is calculated from \refEquation{eq:LippmannSchwinger}, from which we can obtain a solution if $\tau(z)$ satisfies
\begin{equation} \label{EST:eq:tauz_EST}
\tau(z) = \left(\frac{1}{\lambda} - 4 \pi \int_{0}^{\infty} \frac{1}{z- \frac{p^2}{2 \mu}} |g(p)|^2 p^2 \,dp \right)^{-1}.
\end{equation} 
The value of $\lambda$ can be calculated from \refEquation{EST:eq:condition2_v3}, but if we specify the $s$-wave scattering length, we can immediately obtain it from \refEquation{EST:eq:tauz_EST} in the limit $z\rightarrow 0$. Using \refEquation{eq:scattering_length_relation_t0} the resulting expression is
\begin{equation} \label{EST:eq:lambda_EST}
\lambda = \frac{1}{4 \pi^2 \mu} \left( \hbar \frac{|g(0)|^2}{a} - \frac{2}{\pi} \int_0^{\infty} |g(p)|^2 \,dp \right)^{-1}.
\end{equation}
For the square-well potential, the zero-energy $s$-wave scattering state is also an eigenstate of the Hamiltonian corresponding to the separable potential if 
\begin{equation}
g(p) \propto \frac{1}{\bar{p}} \frac{\bar{q}_0 \cos(\bar{q}_0) \sin(\bar{p})-\bar{p} \cos(\bar{p}) \sin(\bar{q}_0)}{\bar{q}_0^2 - \bar{p}^2},
\end{equation}
which follows from \refEquation{EST:eq:Definition_gstates}. The function $\tau(z)$ can then simply be calculated from Eqs.~(\ref{EST:eq:tauz_EST}) and (\ref{EST:eq:lambda_EST}).

In the rest of this paper we do not consider the full EST expansion, but we only consider the single-term EST approximation that reproduces the zero-energy $s$-wave scattering state of the original potential, and we will refer to it as the single-term EST approximation.

\subsection{Comparison of the separable expansions}
\label{ssec:compare_sep_exp}

The methods described above each have some useful properties. First of all, methods I and II share the convenient property that each two-body bound state with angular momentum quantum number $l$ corresponds to only one specific form factor $g_{n l}(p,z)$. This is obvious from the expansion coefficient $\tau_{nl}(z)$ that only has a simple pole exactly at the two-body binding energy of the $n$th dimer state with quantum number $l$. Therefore one can study the effect of these deeper bound states on the weakly bound Efimov trimers by including and excluding the corresponding terms in the expansion of $t_l(p,p',z)$. This statement also applies to method III when one goes beyond the separable approximation. Note, however, that such an expansion also leads to cross terms in the form factors, which make this method potentially more elaborate.

When the potential is approximated by one separable term using method I or II (or the UPA), this approximated potential results in a weakly bound $s$-wave dimer binding energy which is consistent with the full potential $V$. These separable approximations can thus be used for calculating the Efimov states at small positive scattering lengths close to the atom-dimer threshold, and the results of such calculations can be compared with calculations involving the full potential. However, the single-term EST approximation (method III) supports a dimer state whose binding energy deviates from the one corresponding to the potential $V$ at small positive scattering lengths. This can be seen from Fig.~6 of Ref.~\cite{naidon2014physicalorigin} and Fig.~2 of Ref.~\citep{Greene2016softcoreVdW} in which van der Waals potentials are considered. Therefore we will not use this specific single-term EST approximation to study whether the first excited Efimov state crosses the atom-dimer threshold.

We have numerically confirmed that the separable expansions of methods I and II converge to the analytical expression given by \refEquation{ToffmomentumSqW}. The number of terms that are needed to achieve convergence depends on the depth of the well (or equivalently the scattering length and the number of bound states) and the considered energy. As discussed before, the separable expansion of $t_l(p,p',z)$ obtained by using method II converges slowly at large negative energies $z$ below the depth of the well, whereas the convergence is much faster for method I (as well as for the UPE) at these energies. Therefore method I has the best convergence properties. Nevertheless, method II has a computational advantage over method I when performing three-body calculations in which one scans over the scattering length at fixed three-body energy, such as the calculation of the three-body recombination rate. The convergence of the EST method depends on which wave functions at which energies are chosen to be reproduced by the approximated potential. It is difficult to make this choice in general, so that the EST approach lends itself best to yielding a separable approximation for the off-shell components $t_l(p,p',z)$.

\begin{figure}[hbtp]
    \begin{subfigure}
    \centering
    \includegraphics[width=3.4in]{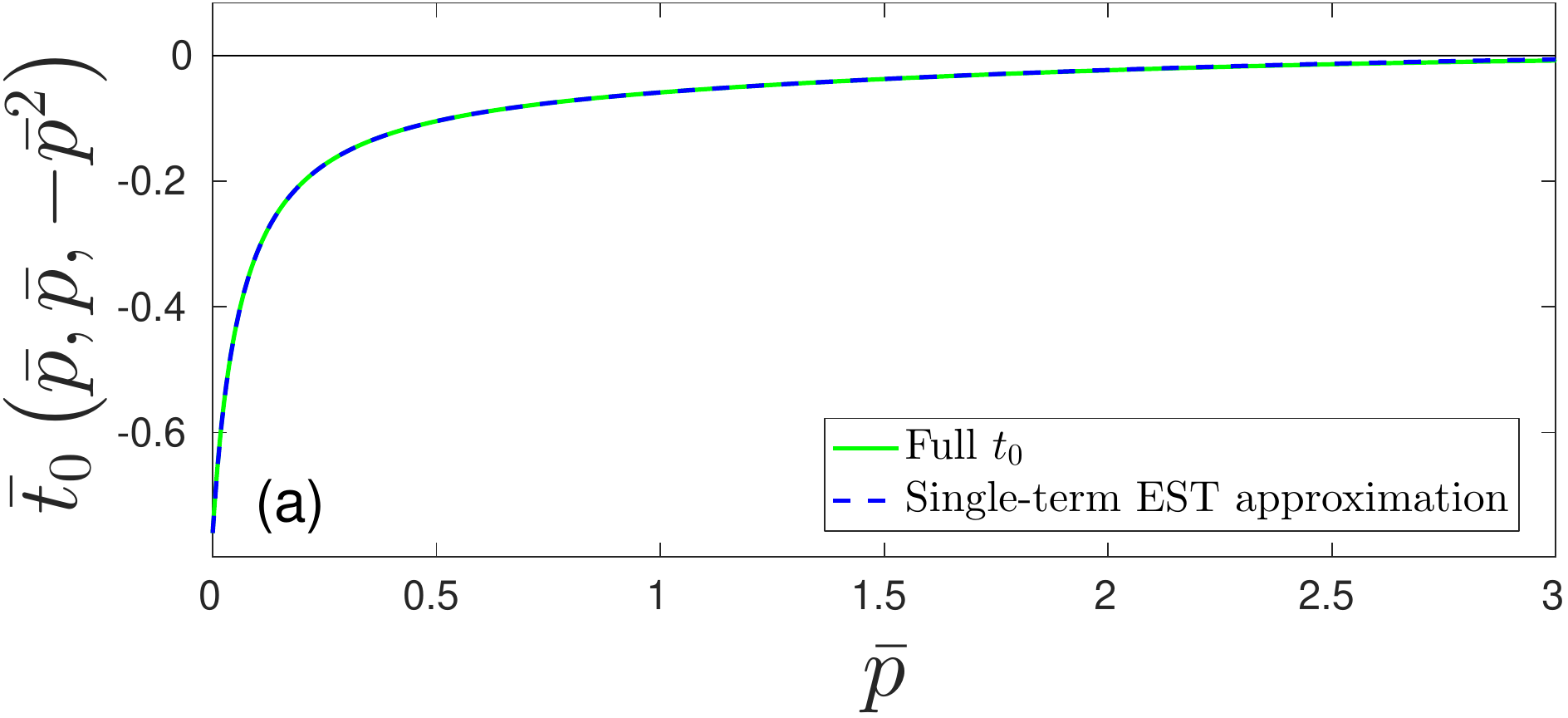}
    \end{subfigure}
    \quad
    
    \begin{subfigure}
    \centering
    \includegraphics[width=3.4in]{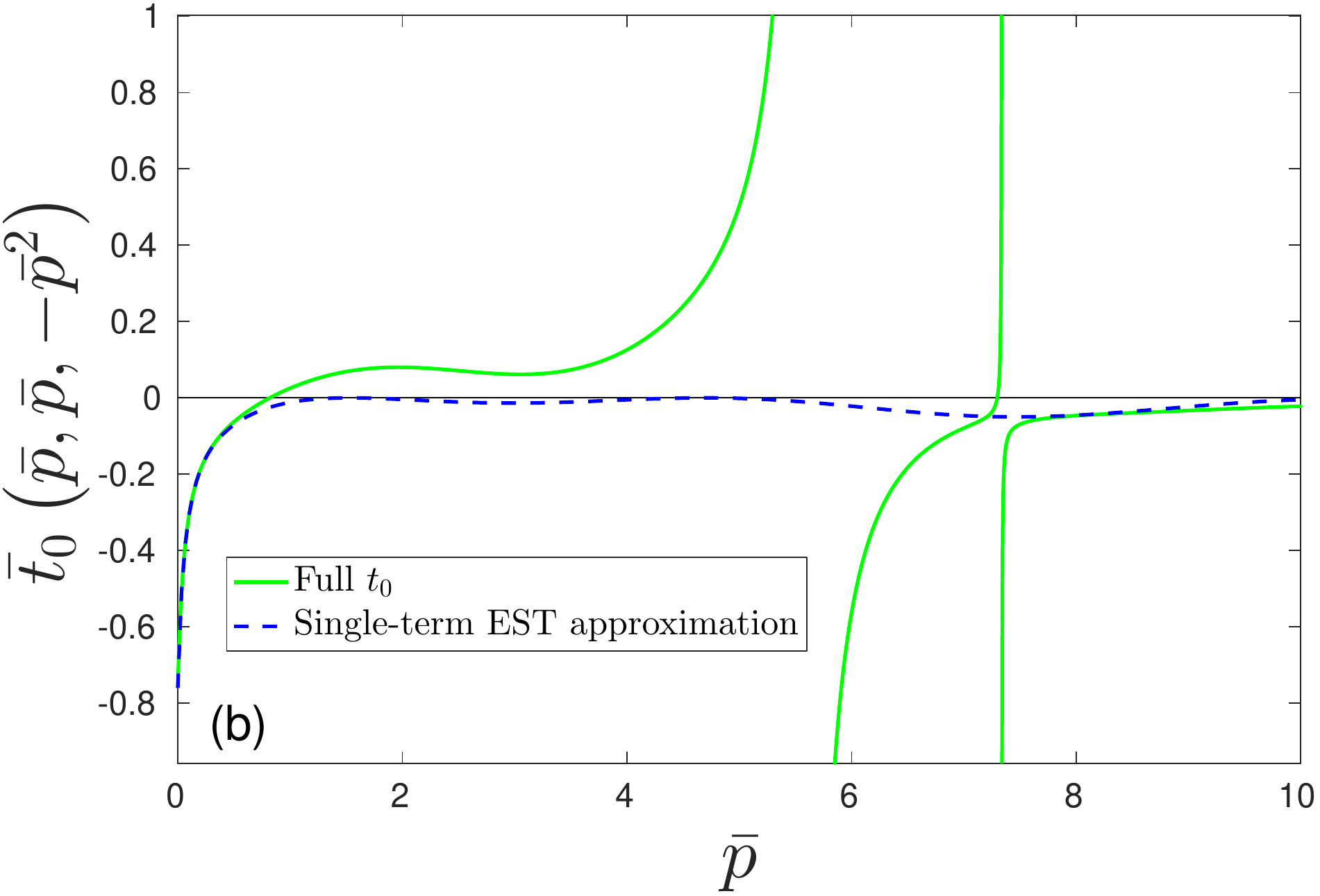}
    \end{subfigure}
    \quad
    
    \caption{Comparison of the off-shell partial-wave component $\bar{t}_0(\bar{p}, \bar{p},-\bar{p}^2) \equiv t_0(p,p,-\frac{p^2}{2 \mu})\cdot m \hbar/R$ corresponding to the square-well potential with the single-term EST approximation (method III). The depth of the square well is chosen such that the first (a) and third (b)  $s$-wave dimer state are almost bound and $\bar{a} = -15$.}
    \label{fig:T2_Comparison_SqW}
\end{figure}

The limitations of the separable approximation become clear when we analyze the approximation for deep square-well potentials. Fig.~\ref{fig:T2_Comparison_SqW} compares the full $s$-wave component $t_0(p,p',z)$ of the square-well potential supporting almost one and three $s$-wave dimer states with the separable EST approximation (method III). The diagonal of $t_0(p,p',z)$, i.e., $p = p'$, is plotted as an example of the behavior of $t_0(p,p',z)$. The figure considers negative energies because these are important in the three-body equations.
Fig.~\ref{fig:T2_Comparison_SqW}(a) shows that the single-term EST approximation  with $\tilde{E}=0$ is a fine substitute for $t_0(p,p',z)$ for the shallow square-well potential. In this case, the potential does not support any bound states, so that no poles are present in $t_0(p,p',z)$ for $z \leq 0$. Deeper potentials support two-body bound states, and this is reflected in the poles of the off-shell $T$ matrix.  Fig.~\ref{fig:T2_Comparison_SqW}(b) shows that the EST approximation works well at small momenta and small negative energies, but it fails for $|\bar{p}_z| \gtrsim 0.5$, where $\bar{p}_z$ is defined in \refAppendixSection{app:Toff_SqW}. The failure of the single-term EST approximation for $|\bar{p}_z| \gtrsim 0.5$ occurs when at least one dimer state is bound. Since the three-body equations involve the off-shell $T$ matrix for all energies $z$ below the considered three-body energy, we expect that the single-term EST approximation gives different results on the three-body level compared to the full square-well potential.

\subsection{Comparison with van der Waals potentials}\label{sec:SepExp:vdW}

Here we compare the form factors between the square-well and a van der Waals potential by using method II (the Weinberg expansion). The considered van der Waals potential is the Lennard-Jones potential $V_{LJ}(r)$ given by
\begin{equation} 
V_{LJ}(r) = - \frac{C_6}{r^6} \left(1 - \frac{\sigma^6}{r^6} \right).
\end{equation}
The dispersion coefficient $C_6$ determines the van der Waals length by $r_{vdW} = (m C_6/\hbar^2)^{1/4}/2$.

\begin{figure}[hbtp]
    \begin{subfigure}
    \centering
    \includegraphics[width=3.4in]{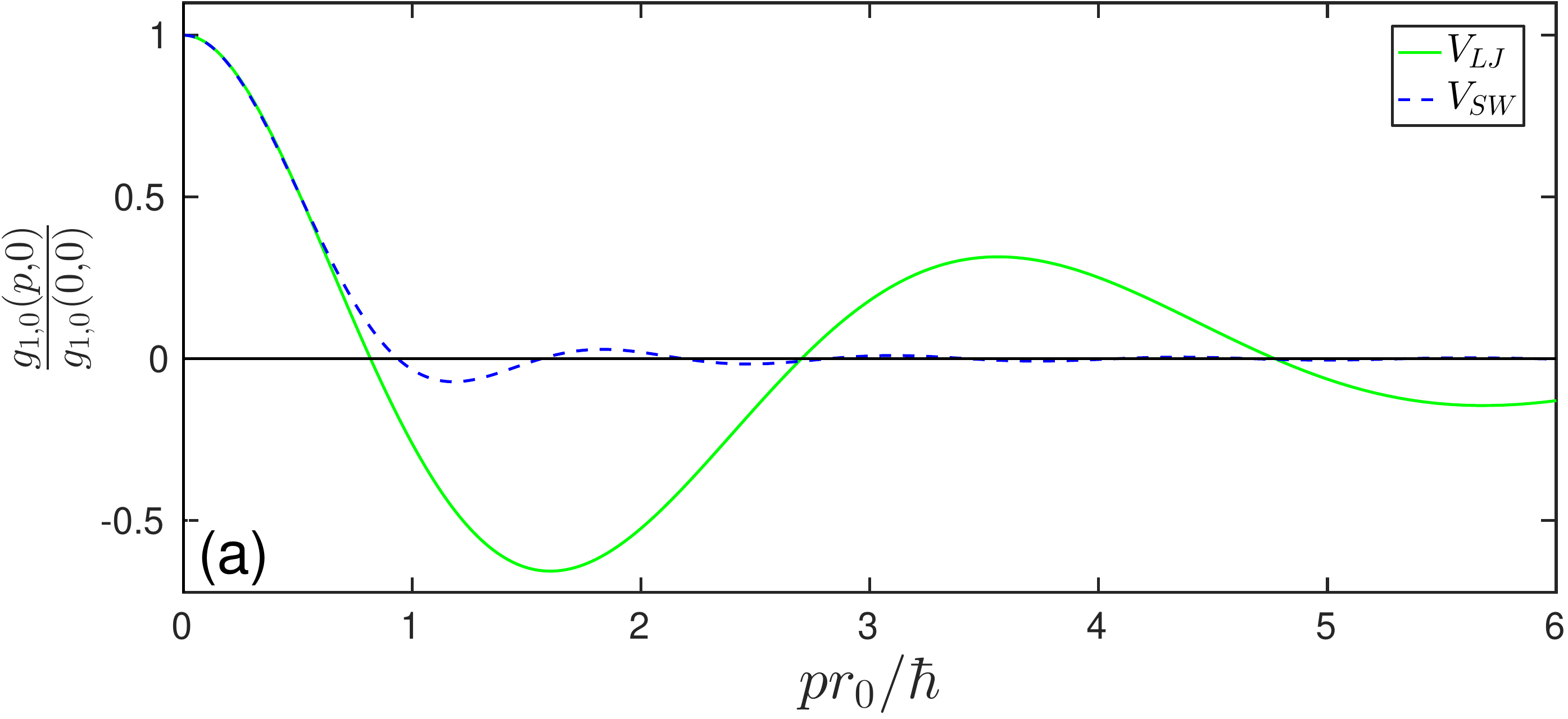}
    \end{subfigure}
    \quad
    
    \begin{subfigure}
    \centering
    \includegraphics[width=3.4in]{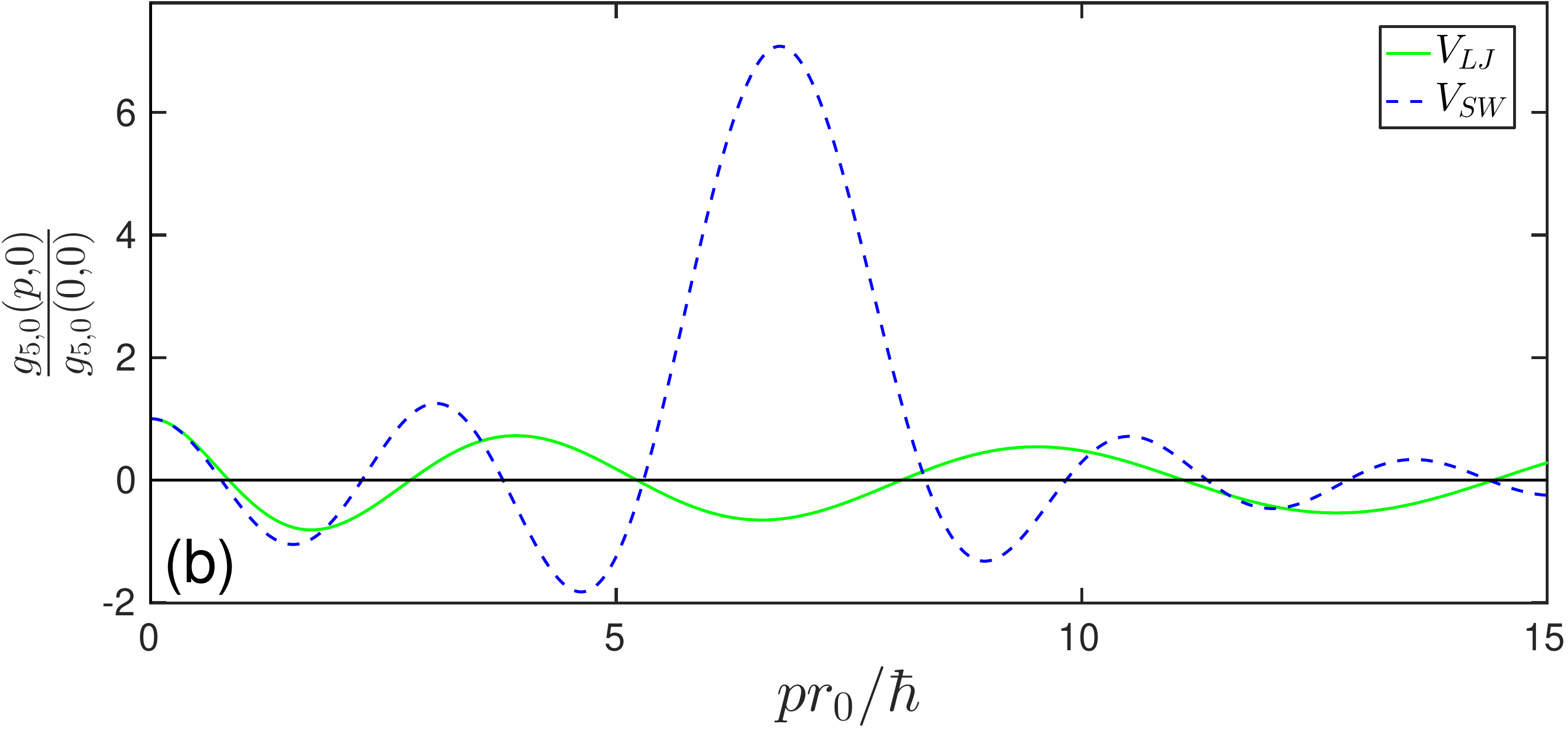}
    \end{subfigure}
    \quad
    
    \caption{Form factors $g_{n l}(p,0)$ of the Weinberg expansion corresponding to the first (a) and fifth (b) $s$-wave two-body bound states of the potentials $V_{LJ}$ and $V_{SW}$, i.e., $l = 0$ and (a) $n = 1$ or (b) $n = 5$. The potentials support exactly one $s$-wave bound state (a) or five $s$-wave bound states (b) and the inverse scattering length $1/a$ is set to zero. We defined the length scale $r_0 = r_{vdW}$ for $V_{LJ}$ and $r_0 = R/x$ for $V_{SW}$ where $x = 5.03$ (a) or $x = 2.08$ (b). These values for $x$ are chosen such that the second derivatives of both form factors match at zero momentum.}
    \label{fig:FormFactors_Comparison_Vlj_Vsw_units_rvdW}
\end{figure}

Fig.~\ref{fig:FormFactors_Comparison_Vlj_Vsw_units_rvdW}(a) compares the form factors obtained for the potentials $V_{LJ}(r)$ and $V_{SW}(r)$ at the first potential resonance (i.e., $z = 0$, $1/a = 0$). The $x$ axis is scaled such that the small-momentum parts of both form factors match well. The main difference occurs at large momenta, where the form factor $g_{1,0}(p,0)$ of the square-well potential drops off to zero much faster than the one of the Lennard-Jones potential. On the three-body level, the small-momentum part is expected to be dominant for the calculation of the Efimov states, especially when the potential does not support any deeper dimer states. General features of the Efimov spectrum of both shallow potentials can thus be compared and we will use this result in \refSection{ssec:Results:shallow}.

For potentials supporting more dimer states, the situation changes significantly. Fig.~\ref{fig:FormFactors_Comparison_Vlj_Vsw_units_rvdW}(b) compares the fifth form factor obtained for the potentials $V_{LJ}(r)$ and $V_{SW}(r)$ at the fifth potential resonance. The small-momentum part can again be matched reasonably well by scaling the $x$ axis. However, the large-momentum part of both form factors behaves in a completely different way. The form factor $g_{5, 0}(p,0)$ of the square-well potential shows a large peak which does not occur in the form factor of the Lennard-Jones potential. In general, the large peak of the square-well form factor $g_{n, 0}(p,0)$ occurs near $\bar{p}\simeq (2 n -1) \pi/2$ (except for $n = 1$; see Fig.~\ref{fig:FormFactors123_VG0_Comparison_q2_0}). The energy corresponding to this momentum $\bar{p}$ is the depth of the square well for which the $n$th $s$-wave dimer state occurs at energy $z = 0$. The huge peak arises from the fact that the zero-energy $s$-wave scattering wave function is not suppressed inside the well as was pointed out by Naidon \textit{et al.} \cite{naidon2014microscopic}. The presence of the large peak is therefore a feature of the square-well potential, and it is present in the form factors of all three considered expansion methods discussed above. As a result, the large-momentum part of the two-body $T$ matrix is expected to be non-negligible for calculating the Efimov states of deep square-well potentials, which we will indeed observe in \refSection{sec:Results:Deep}.

\section{Three-body properties including finite-range effects}\label{sec:Results}

Now we focus on the Efimov states corresponding to the potential resonances of the square-well potential. First we consider a shallow square-well potential supporting only one $s$-wave dimer state, and analyze the corresponding Efimov spectrum. In particular, we focus on the behavior of the Efimov states near the atom-dimer threshold and investigate the effects of $d$-wave interactions on the three-body level. After having analyzed the full Efimov spectrum, we investigate how good a separable approximation for the square-well potential is for the determination of the Efimov spectrum. Additionally, we discuss the convergence of the different methods described above to expand the off-shell $T$ matrix \cite{noteCalcScatLength}.
In the second part, we consider deeper square-well potentials and investigate the validity of the separable approximation for calculating the three-body parameter.
For the second and third potential resonance of the square-well potential, we have included all partial-wave components that are necessary to obtain converged results, and analyze the behavior of the Efimov states near the atom-dimer threshold.

\subsection{Shallow square well}\label{ssec:Results:shallow}

\begin{figure}[hbtp]
    \centering
    \includegraphics[width=3.4in]{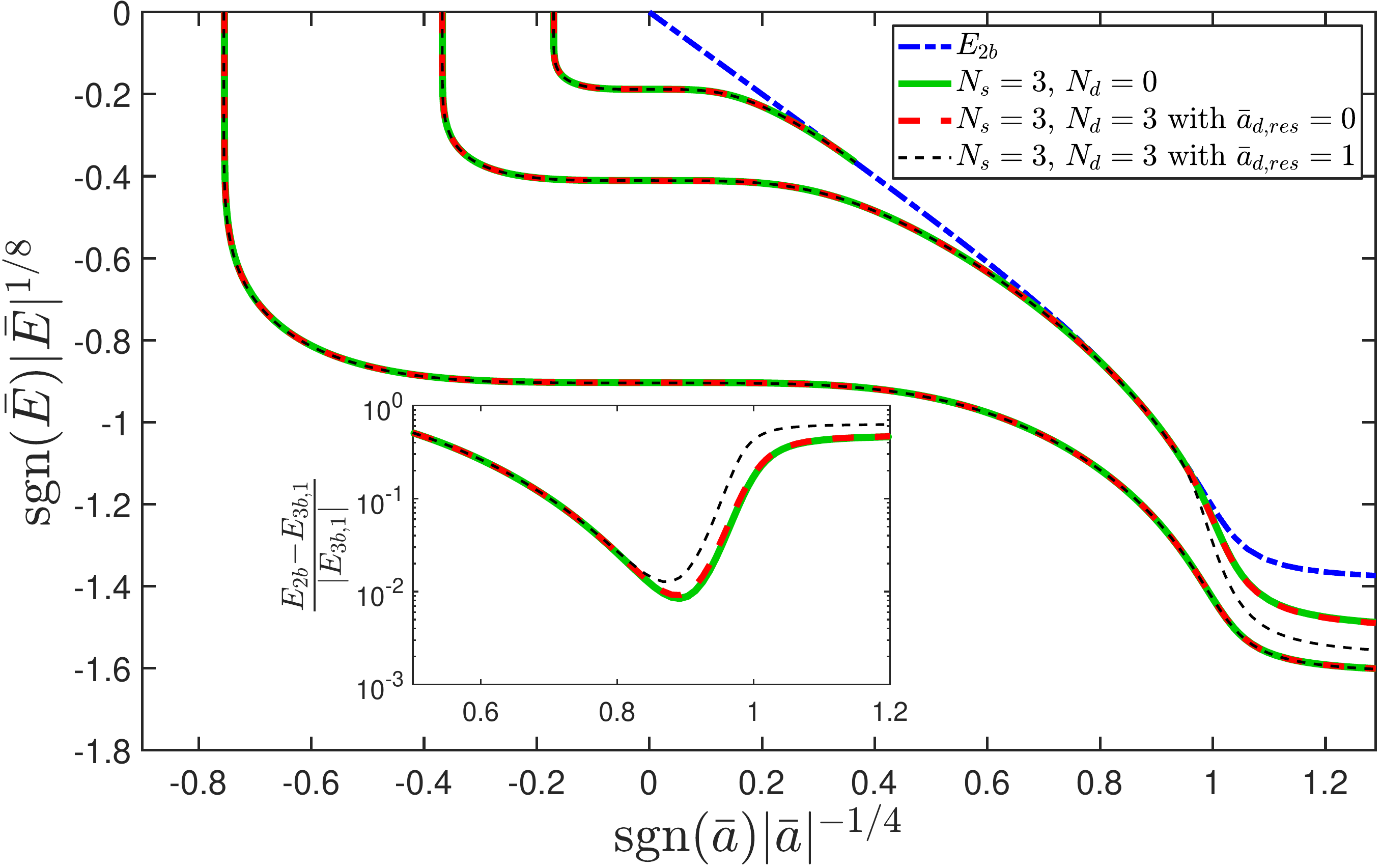}
    \caption{Energy of the lowest three Efimov states calculated near the first potential resonance of the square-well potential by using method I for $N_s = 3$ and $N_d = 0$ and for $N_s = 3$ and $N_d = 3$, where the numbers $N_s$ and $N_d$ indicate the number of separable terms that are used to approximate the partial-wave components $t_l(p,p',z)$ for $l = 0$ and $l = 2$, respectively. The black dashed curves indicate the calculation for which the $d$-wave resonance is artificially changed from $\bar{a} = 0$ to $\bar{a} = 1$. The blue line is the binding energy corresponding to the $s$-wave dimer state. The inset shows the relative energy difference between the energies of the $s$-wave dimer state and the first excited Efimov state as a function of the inverse scattering length.}
    \label{fig:PotRes1_Spectrum_3S3D_3S0D}
\end{figure}

Here we focus on the first potential resonance of the square-well potential. Fig.~\ref{fig:PotRes1_Spectrum_3S3D_3S0D} shows the energies of the lowest three Efimov states as a function of the inverse $s$-wave scattering length near this potential resonance. The corresponding three-body parameters are given in Table~\ref{tab:3BP_SqW_PotRes1} in \refAppendixSection{app:tables}. Fig.~\ref{fig:PotRes1_Spectrum_3S3D_3S0D} shows that not only the ground Efimov state does not cross the atom-dimer threshold, but also the first excited Efimov state stays below this threshold.
This can be seen from the inset in Fig.~\ref{fig:PotRes1_Spectrum_3S3D_3S0D} in which the relative energy difference between the energies of the $s$-wave dimer state and the first excited Efimov state is shown as a function of the inverse scattering length. The noncrossing of the ground Efimov state with the atom-dimer threshold is also found for shallow van der Waals potentials \cite{mestrom2017jila,Greene2016softcoreVdW,cong2016vdWaals} and is consistent with a variational principle \cite{bruch1973variational} that constrains the ground-state energy of three identical bosons, interacting via spherically symmetric pair potentials such as the square-well potential, to always lie below the ground-state energy of two of such bosons, more precisely $|E_{3b,0}|\geq 3 |E_{2b,0}|$. 

\begin{figure}[hbtp]
    \centering
    \includegraphics[width=3.4in]{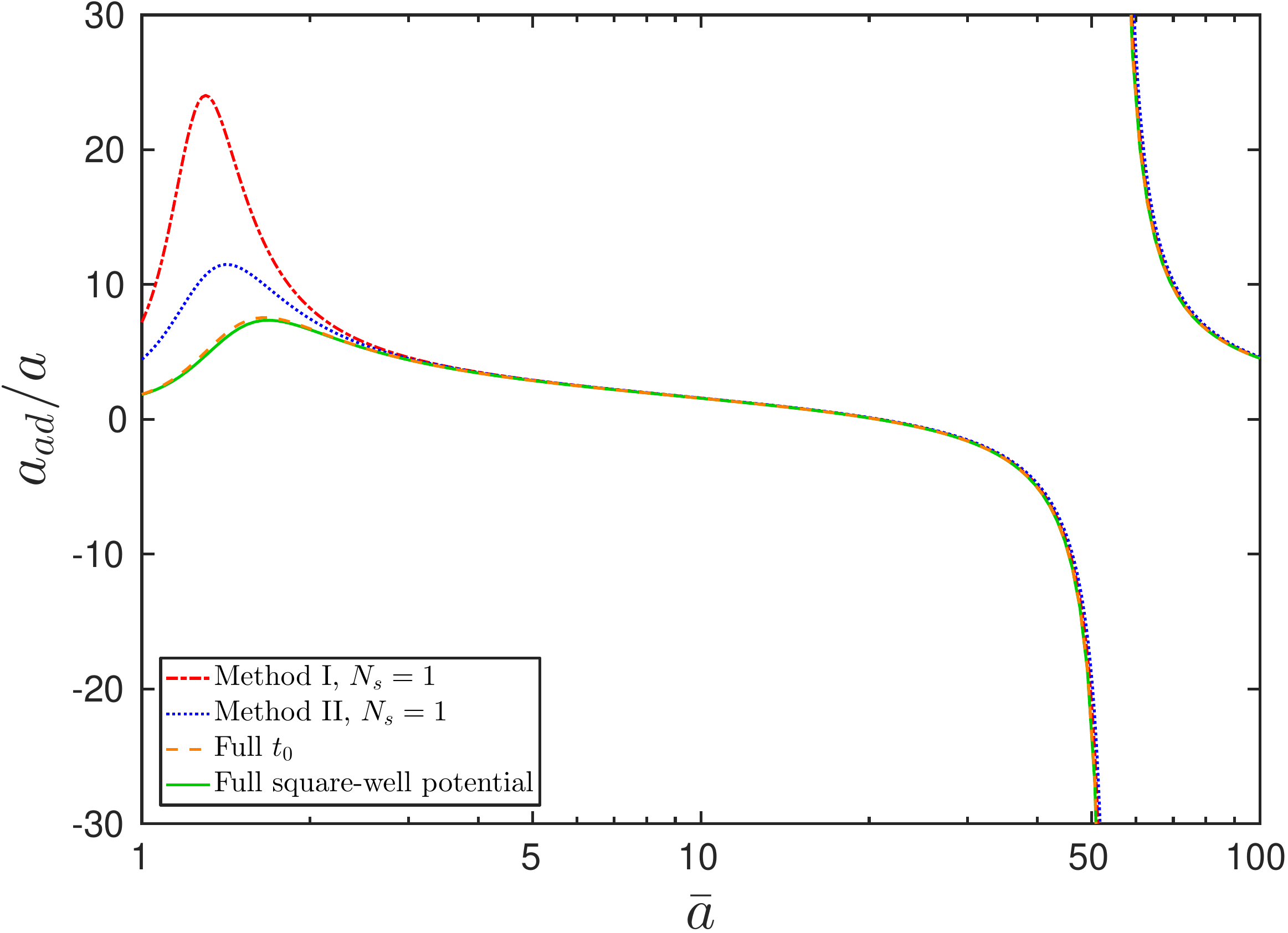}
    \caption{The $s$-wave atom-dimer scattering length as a function of the $s$-wave two-body scattering length $a$ for the shallow square-well potential and several approximations. The red dashed-dotted curve and the blue dotted curve involve separable approximations of $t_0(p,p',z)$, whereas the orange dashed curve results from three-body calculations involving the full $s$-wave component of the $T$ matrix and has been obtained using both method I and method II. The green curve corresponds to calculations using method II and involves both the $s$-wave and $d$-wave components of the off-shell $T$ matrix. We have confirmed that the inclusion of $t_4(p,p',z)$ hardly changes the atom-dimer scattering length.}
    \label{fig:PotRes1_AD}
\end{figure}

The first excited Efimov state corresponding to the shallow square-well potential does not merge with the atom-dimer threshold. This is also reflected in the $s$-wave atom-dimer scattering length shown in Fig.~\ref{fig:PotRes1_AD}. This figure shows only one atom-dimer resonance that occurs at $\bar{a}_{*,2} = 54.5$ and corresponds to the crossing of the second excited Efimov state with the atom-dimer threshold.
Another interesting feature occurs at small positive scattering lengths. The value of $a_{ad}$ shows a maximum at $\bar{a} = 2.07$, but it does not diverge. This indicates that the first excited Efimov state approaches the atom-dimer threshold closely for decreasing $a$, but it does not become unbound. As $a$ decreases further, the binding energy of this trimer, $E_b = E_{2b}-E_{3b}$, state increases. 

The noncrossing of the first excited Efimov state has also been seen before for the Lennard-Jones potential in Ref.~\cite{mestrom2017jila} in which this effect was attributed to strong $d$-wave interactions for van der Waals potentials near $a = 1$~$r_{vdW}$ \cite{dincao2012dwave,mestrom2017jila}. This hypothesis was not confirmed because the $d$-wave interactions cannot be excluded in the adiabatic hyperspherical representation used by Ref.~\cite{mestrom2017jila}. Our method allows us to include or exclude $d$-wave interactions. Fig.~\ref{fig:PotRes1_Spectrum_3S3D_3S0D} additionally compares the calculation in which only $s$-wave effects are included versus one in which both $s$-wave and $d$-wave interactions are taken into account. The resulting curves clearly overlap from which we conclude that the effect of the $d$-wave interactions on the Efimov states is small for this shallow square-well potential. This is not surprising since the $d$-wave dimer becomes bound at $\bar{a} = 0$. For single-channel interactions with a van der Waals tail, $-C_6 r^{-6}$, the $d$-wave dimer always becomes bound at a scattering length $a = 4 \pi/[\Gamma(1/4)]^2 \approx 0.956$ $r_{vdW}$ as predicted by Gao \cite{gao2000waals}. This prediction has been confirmed by Wang \textit{et al.} \cite{dincao2012dwave} using the Lennard-Jones potential as a two-body interaction.

In order to investigate the effect of strong $d$-wave interactions that are present for van der Waals potentials at small positive scattering lengths, we artificially increase the strength of the $d$-wave interactions by making the depth $V_0$ of the square well larger for the $d$-wave partial-wave component $t_2(p,p',z)$ in the three-body calculation. In this way the $d$-wave resonance is closer to the $s$-wave resonance. Fig.~\ref{fig:PotRes1_Spectrum_3S3D_3S0D} also compares the energies of the Efimov states for calculations involving the weak (unmodified) $d$-wave interactions and the strong (modified) $d$-wave interactions in which the $d$-wave resonance occurs at $\bar{a} = 1$. The increase of the $d$-wave interaction strength has almost no effect on the ground Efimov state because the $d$-wave dimer state is well separated in energy. However, the first excited Efimov state is strongly affected at small positive scattering lengths, where the energy of this trimer state is decreased. So indeed strong $d$-wave effects can be the cause of the noncrossing of the first excited Efimov state with the two-body threshold for the potential resonances of the Lennard-Jones potential as seen in Ref.~\cite{mestrom2017jila}.

\begin{figure}[hbtp]
    \centering
    \includegraphics[width=3.4in]{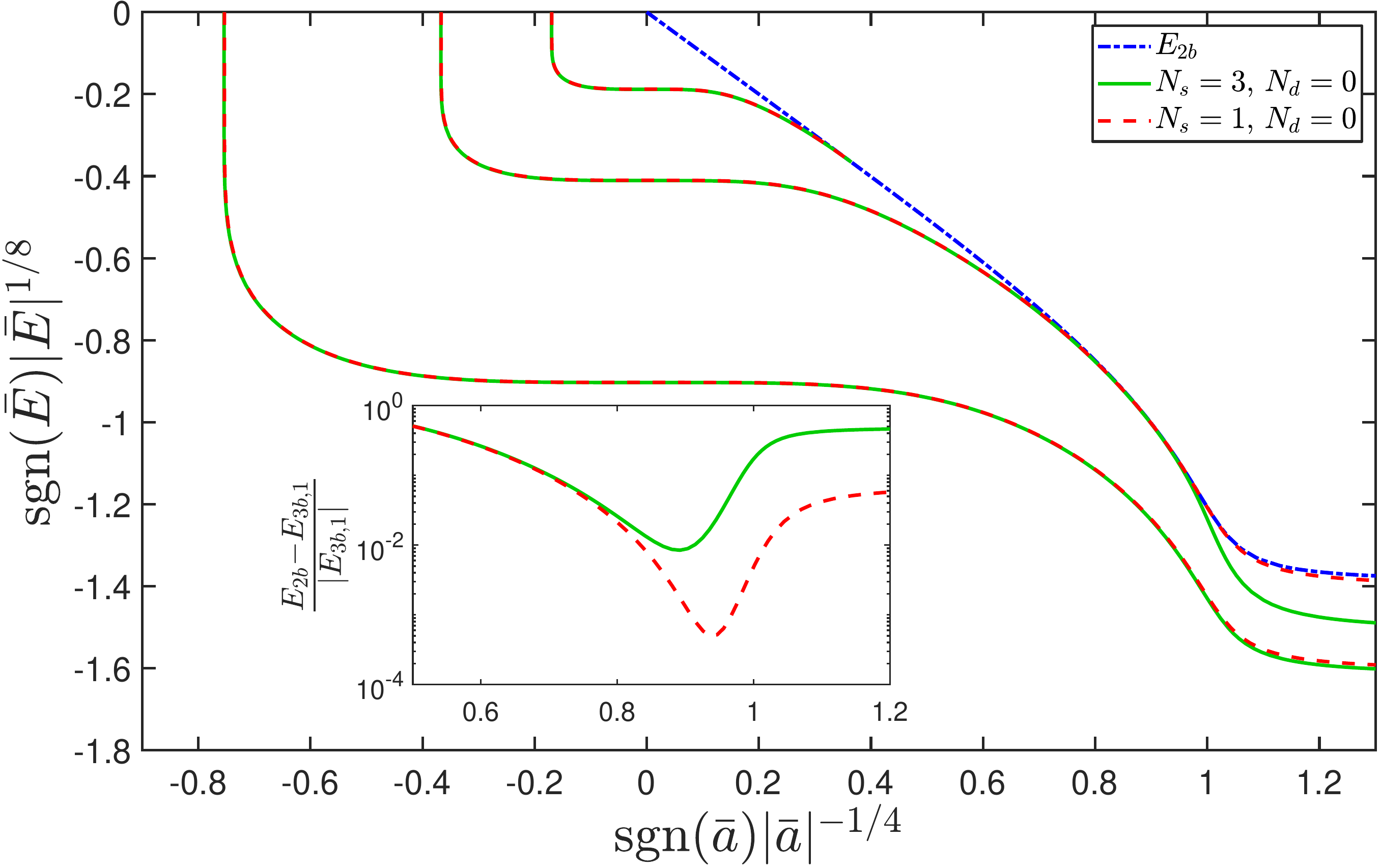}
    \caption{Energy of the lowest three Efimov states calculated near the first potential resonance of the square-well potential by using method I for $N_s = 3$ and $N_s = 1$. In both cases $N_d$ is set to zero. The blue line is the binding energy corresponding to the $s$-wave dimer state. The inset shows the relative energy difference between the energies of the $s$-wave dimer state and the first excited Efimov state as a function of the inverse scattering length \cite{repeated3bodyCalc}.}
    \label{fig:PotRes1_Spectrum_3S0D_1S0D}
\end{figure}

The way which has so far been used most for calculating the energies of the Efimov states via the Faddeev equations is to approximate the $s$-wave component of the two-body $T$ matrix by one separable term, neglecting all higher-order partial-wave components, and to solve the resulting integral equation. This method is believed to work well because the nonseparable function $t_0(p,p',z)$ is very separable in the energy regime in which the Efimov states are located. After all, the $s$-wave component $t_0(p,p',z)$ is more separable for energies $z$ closer to the energy of a two-body $s$-wave bound state \cite{faddeev1993scattering}. In Fig.~\ref{fig:PotRes1_Spectrum_3S0D_1S0D} we compare the Efimov spectrum corresponding to the full $s$-wave component of the $T$ matrix with the one corresponding to its single-term approximation. This figure shows that the separable approximation works reasonably well, except for the first excited Efimov trimer at small positive scattering lengths, which is strongly affected by the remaining terms of the two-body $T$ matrix. In both cases the first excited Efimov state does not cross the atom-dimer threshold as can be seen from the inset, but it stays much closer to the two-body threshold at small positive scattering lengths when the off-shell $T$ matrix is approximated by a fully separable function. The same conclusion follows from Fig.~\ref{fig:PotRes1_AD} in which the atom-dimer scattering length of the shallow square-well potential is compared with the single-term approximations of methods I and II. The reason why the use of the separable approximation fails at large negative energies close to the dimer threshold is not obvious as the off-shell $T$ matrix is highly separable in this regime. Instead, the cause of this failure is related to the Green's function $G_0$ that is present in the Faddeev equations, \refEquation{OperatorPhi}. This  Green's function is represented in \refEquation{3bodyCoupledEq} by the factor $1/\left(E-\left(q^2+ \mathbf{q}\cdot \mathbf{q}' + q'^2 \right)/m\right)$. The factor $q'^2 /\left(E-\left(q^2+ \mathbf{q}\cdot \mathbf{q}' + q'^2 \right)/m\right)$ in \refEquation{3bodyCoupledEq} clearly suppresses the small-momentum part, i.e., $q' \ll \hbar/R$, in which $\tau_{1,0}\left(E-3 q'^2/(4 m) \right)$ is the biggest. When the three-body energy $E$ is not close to zero, this suppression is much more effective. 
As a result, the dominance of the first term in the separable expansion is reduced in the determination of the three-body bound states and the separable approximation for $t_0(p,p',z)$ is not sufficient to calculate the first excited Efimov state accurately at energies roughly below  $-\hbar^2 /(2 \mu R^2)$.

Based on this reasoning, one would expect that the separable approximation would also fail in the calculation of the ground Efimov state at large negative energies. However, the energy of the ground Efimov state at small positive scattering lengths is quite similar for both calculations shown in Fig.~\ref{fig:PotRes1_Spectrum_3S0D_1S0D}. We attribute this effect to the variational principle \cite{bruch1973variational} stating that $|E_{3b,0}|\geq 3 |E_{2b,0}|$, so that the binding energy of the ground Efimov state, which is close to this limiting value, cannot decrease much at small positive scattering lengths for a decreasing number of terms in the separable expansion. Even though this variational principle is proven for energy-independent potentials \cite{bruch1973variational, lim1979raregas}, its proof is based on the two-body ground state wave function of two identical spinless bosons interacting via a spherically symmetric potential, and thus holds for both the square-well potential and its separable approximation obtained by the spectral representation (method I) or the Weinberg expansion (method II).

Table~\ref{tab:3BP_SqW_PotRes1} in \refAppendixSection{app:tables} summarizes the three-body parameters near the first pole of the $s$-wave scattering length calculated from methods I, II and III. The wave number $\kappa_n^*$ corresponds to the energy $E_n^* = -\left(\hbar \kappa_n^*\right)^2/(2 \mu)$ of the $n$th trimer state at diverging scattering length. The three-body parameters calculated from method I converge the fastest as more expansion terms are included. The results of method II converge less fast because the form factors do not depend on the scattering length for fixed range $R$. Furthermore, method I provides the best single-term approximation, followed by method III and II respectively. This result is not expected to hold in general, but only for the first potential resonance. The EST approximation is expected to be the best single-term approximation for deeper potentials because it reproduces the correct zero-energy two-body scattering state. Table~\ref{tab:3BP_SqW_PotRes1} also shows that the relative difference between the calculations with and without $d$-wave effects is smaller than $10^{-3}$, so that $d$-wave effects might need to be considered depending on the required accuracy. This result only holds for the first potential resonance. The $d$-wave effects are larger for deeper potentials that also support $d$-wave dimer states as is described in the next section.

\subsection{Deeper square wells} \label{sec:Results:Deep}

For the shallow square-well potential, we have found that a separable approximation for the off-shell $T$ matrix works quite well in order to determine the three-body parameter. The validity of the separable approximation for deeper square-well potentials is nontrivial since we have concluded from \refSection{ssec:compare_sep_exp} that the energy dependence of the off-shell $T$ matrix of square-well potentials supporting more than one two-body bound state is not correctly approximated by using the single-term EST approximation. For other classes of finite-range interactions (including van der Waals potentials), it has been shown that the single-term EST approximation seems to give reasonable results for the three-body parameter \cite{naidon2014microscopic}. The square-well potential does not belong to one of these classes. Therefore we go beyond the separable approximation to test its validity.

\begin{figure}[ht]
    \centering
    \includegraphics[width=3.4in]{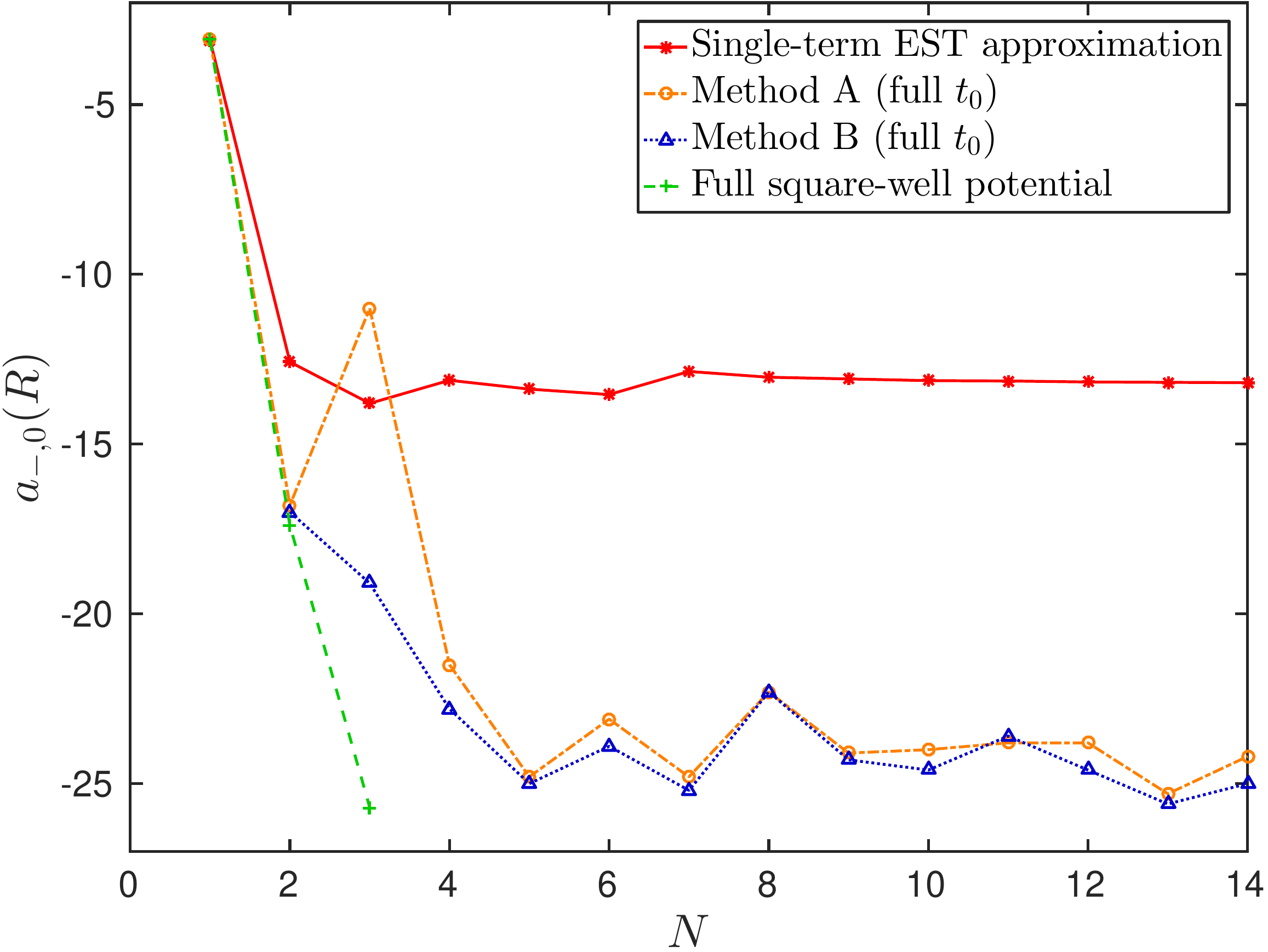}
    \caption{Values for the three-body parameter $\bar{a}_{-,0}$ corresponding to the $N$th potential resonance of the square-well potential. The red data points (*) are obtained using the single-term EST approximation, whereas the orange ($\circ$) and blue ($\bigtriangleup$) data points involve the full expansion of the $s$-wave component $t_0(p,p',z)$ using the Weinberg expansion. The green ($+$) data points correspond to the full square-well potential. The corresponding data can be found in Tables~\ref{tab:3BP_SqW_methodEST}, \ref{tab:3BP_SqW_methodVG0_l_0} and \ref{tab:3BP_SqW_methodVG0_l_total}.}
    \label{fig:3BP_deep_SqW_VG0_vs_EST}
\end{figure}

First, we consider separable $T$ matrices based on the EST method. The corresponding results are shown in Fig.~\ref{fig:3BP_deep_SqW_VG0_vs_EST} and in Table~\ref{tab:3BP_SqW_methodEST} in \refAppendixSection{app:tables} in which the three-body parameters are given as a function of the $N$th potential resonance. The figure shows that the three-body parameter of the square well converges as a function of the number of two-body $s$-wave bound states when using the single-term EST approximation. Interestingly, it shows a large jump between the three-body parameters for $N = 1$ and $N > 1$. This is a typical feature for the square-well potential, which is caused by a quite distinct shape of the form factor $g_{1,0}(p)$ compared to the other form factors $g_{n,0}(p)$ for $n>1$, that are quite similar for small momenta $p$.

The single-term EST approximation for $t_0(p,p',z)$ corresponds to a potential that does not support any deeper two-body bound states, so that the corresponding Efimov states are true bound states. The deeper molecular states can be included by using the Weinberg expansion to approximate the two-body $T$ matrix. Considering only $s$-wave interactions, we include as many terms as are necessary to get converged results for the three-body parameter $a_{-,0}$ using the two methods described in \refSection{sec:FaddeevResonances} and \refSection{sec:K3} (indicated by A and B, respectively). These results are presented in Fig.~\ref{fig:3BP_deep_SqW_VG0_vs_EST} and in Table~\ref{tab:3BP_SqW_methodVG0_l_0} in \refAppendixSection{app:tables}. They differ completely from the results of Table~\ref{tab:3BP_SqW_methodEST} that were obtained upon using the single-term EST approximation, even though both methods reproduce the same $s$-wave component of the two-body scattering wave function at zero energy. This difference is caused by the fact that this separable approximation is not a good substitute for the $s$-wave component of the two-body $T$ matrix at larger negative energies that are still relevant for the three-body calculations as discussed in \refSection{ssec:compare_sep_exp}. Fig.~\ref{fig:3BP_deep_SqW_VG0_vs_EST} also shows that the full expansion of $t_0(p,p',z)$ leads to a three-body parameter $a_{-,0}$ of which the convergence as a function of the depth of the square well is less fast than for the single-term EST approximation. We attribute this effect to the large peak of the form factors of the square-well potential (see \refSection{sec:SepExp:vdW}) that is present in every expansion term of the function $t_0(p,p',z)$.

The standard approach to determine $a_{-}$ is to locate the maxima in the three-body recombination rate (method B). Table~\ref{tab:3BP_SqW_methodVG0_l_0} shows that the results obtained by searching for which scattering length the real part of the relevant eigenvalue (indicated by $\varepsilon$) equals one (method A) agrees quite well with method B when the inelasticity parameter $\eta_{*}$ is small in which case the imaginary part of $\varepsilon$ is also small. Even though method A is not suitable for determining the Efimov states at high accuracy, it can be used to get a rough estimate. Furthermore, method A can also be used at negative energies, so that it could be used to find the full Efimov spectrum.

So far we have only included $s$-wave bound states, but other dimer states with even angular momentum quantum number $l$ should be important as well for the square-well potential. They enter the three-body equation via the partial-wave components $t_l(p,p',z)$ which we again expand in separable terms using the Weinberg expansion. The results of these calculations for the second and third potential resonance are shown in \refFigure{fig:3BP_deep_SqW_VG0_vs_EST} and in Table~\ref{tab:3BP_SqW_methodVG0_l_total} in \refAppendixSection{app:tables}. The three-body parameter for second and third potential resonance is $\bar{a}_{-,0} = -17.4 \pm 0.1$ and $\bar{a}_{-,0} = -25.7 \pm 0.1$ respectively. The Efimov resonance is thus indeed (strongly) affected by the higher angular momentum components. At negative scattering lengths near the second potential resonance ($N = 2$), the square-well potential only supports two dimer states: one $s$-wave and one $d$-wave dimer state. Thus the Efimov state is pushed upward by this $d$-wave two-body bound state. Similarly, the Efimov state near the third potential resonance also shifts upward by the inclusion of the dimer states with quantum numbers $l = 2$ and $l = 4$. The results of Table~\ref{tab:3BP_SqW_methodVG0_l_total} show that it is insufficient to include only $s$-wave interactions in the calculation of the Efimov resonances. The higher partial-wave components labeled by $l$ become important when the two-body potential supports bound states with angular momentum quantum number $l$. These effects are not included in the single-term EST approximation, so that the three-body parameter calculated upon using the single-term EST approximation deviates strongly from the actual three-body parameter for $N = 2$ and $N = 3$. 

\begin{figure}[hbtp]
    \centering
    \includegraphics[width=3.4in]{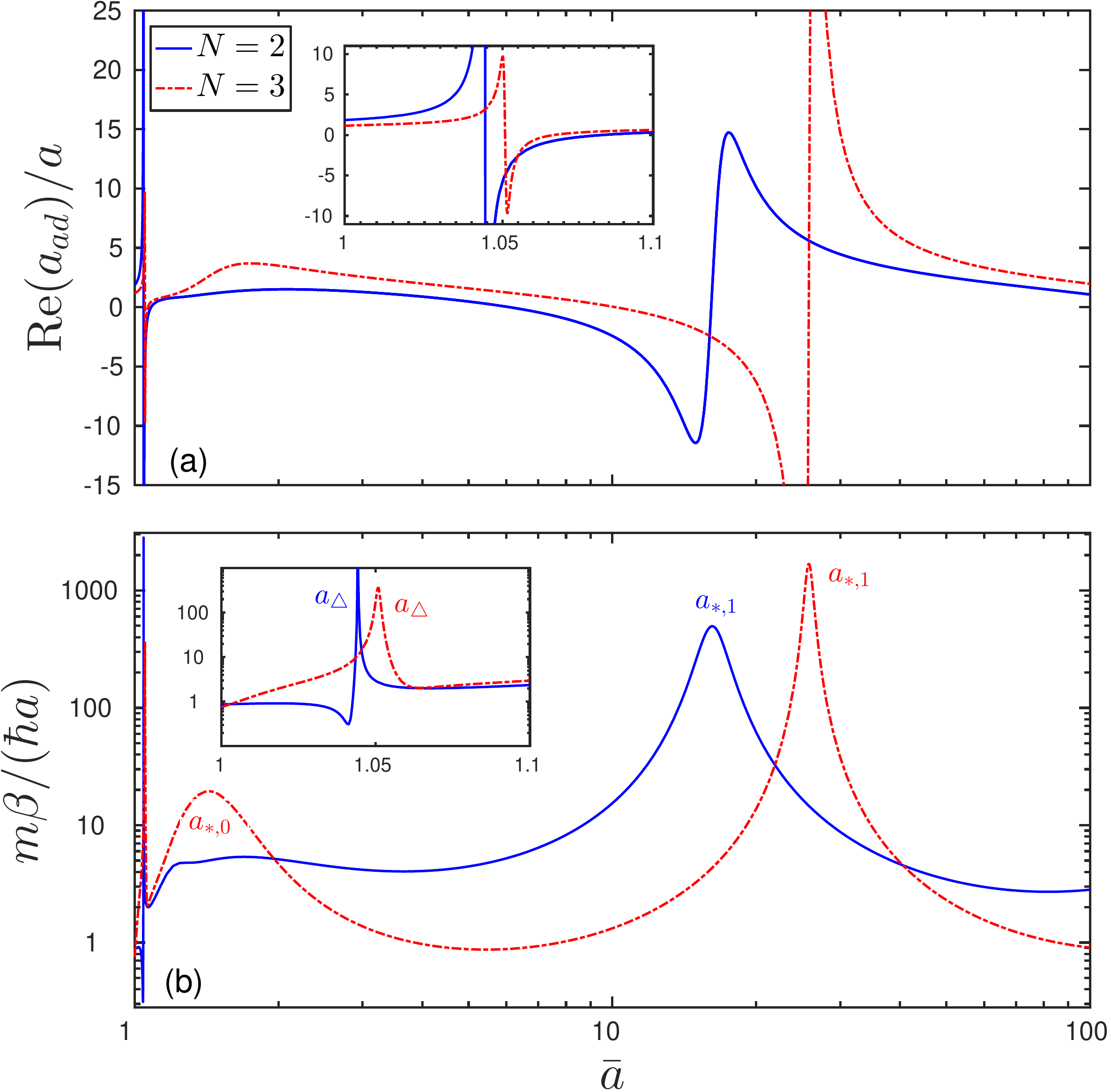}
    \caption{(a) The atom-dimer scattering length $a_{ad}$ and (b) the corresponding inelastic scattering rate for positive scattering lengths near the $N$th potential resonance of the square-well potential. The $T$ matrix is expanded using the Weinberg expansion and the partial-wave components with $l = 0, 2, 4, 6$, and $8$ are taken into account. The inset zooms in on small positive scattering lengths and displays resonant behavior due to trimer resonances.}
    \label{fig:PotRes_2_and_3_AD_and_beta}
\end{figure}

Until now we have only discussed the three-body parameter $a_{-,0}$ for square-well potentials supporting several bound states. Now we turn our attention to positive scattering lengths. Figure \ref{fig:PotRes_2_and_3_AD_and_beta} shows the real part of the atom-dimer scattering length and the atom-dimer loss rate for positive scattering lengths near the second and third potential resonance of the square-well potential. The considered dimer state in this scattering process is the (weakly bound) $s$-wave dimer state that corresponds to the considered potential resonance. The values at which the loss rate $\beta$ peaks are summarized in \refTable{tab:3BP_SqW_PotRes_2_and_3_astar}. Fig. \ref{fig:PotRes_2_and_3_AD_and_beta}(a) shows that first excited Efimov state merges with the atom-dimer threshold for both potential resonances. This trimer resonance enhances the inelastic atom-dimer scattering cross section as can be seen from Fig. \ref{fig:PotRes_2_and_3_AD_and_beta}(b). We know that this trimer resonance is related to the first excited Efimov trimer due to the universal scaling relations between $a_{-}$ and $a_{*}$ \cite{braaten2006universality}. For the third potential resonance, we see an additional peak in $\beta$ near $\bar{a} = 1.46$ even though the real part of $a_{ad}$ does not change sign. This suggests that the lowest Efimov trimer is close to the atom-dimer threshold at $\bar{a} = 1.46$, but it does not become unbound. 

Furthermore, the square-well potential shows some interesting behavior at small positive scattering lengths. The insets in Fig. \ref{fig:PotRes_2_and_3_AD_and_beta} shows an additional trimer resonance near $\bar{a} = \bar{a}_{\triangle}$ for both the second and third potential resonance. As the scattering length decreases, the real part of $a_{ad}$ changes from negative to positive values at $\bar{a} = \bar{a}_{\triangle}$, which means that an additional trimer state is being formed as the depth of the potential is increased. This trimer state is not an Efimov trimer and it is therefore strongly dependent on the short-range details of the interaction potential.

\begin{table}[h]
  \centering
  \caption{Values of the scattering lengths at which the atom-dimer inelastic scattering rate $\beta$ peaks (see Fig. \ref{fig:PotRes_2_and_3_AD_and_beta}). Positive scattering lengths near the $N$th potential resonance of the square-well potential are considered. The value of $\bar{a}_{*,1}$ is obtained by fitting the data with \refEquation{eq:aAD_Universal}. The loss parameter $\eta_{*}$ resulting from this fit is also indicated.}
  \label{tab:3BP_SqW_PotRes_2_and_3_astar}
  \begin{ruledtabular}
    \begin{tabular}{ccccc}
    \toprule
    $N$ & $\bar{a}_{*,0}$  & $\bar{a}_{*,1}$ & $\eta_{*}$ & $\bar{a}_{\triangle}$ \\
    \hline
    2 & - & 16.2 & 0.08 & 1.044 \\
    3 & 1.46 & 25.8 & 0.02 & 1.051 \\
    \bottomrule
    \end{tabular}\\
    \end{ruledtabular}
\end{table}

\section{Conclusion}\label{sec:Conclusion}

We have studied Efimov physics for a three-body system of identical bosons interacting via a pairwise square-well potential, we analyzed the regime of validity of the corresponding separable potential approximation, and investigated what is needed to improve on this approximation.
For this purpose, we solved the Faddeev equations in the momentum-space representation. These equations depend on the two-body potential via the off-shell $T$ matrix that is nonseparable whenever the considered potential itself is nonseparable. Since the off-shell $T$ matrix of the square-well potential is nonseparable, we expanded this $T$ matrix in separable terms for solving the three-body equations. We described three distinct expansion methods, namely the spectral representation, the Weinberg expansion and the EST expansion, and discussed the advantages and disadvantages of these methods on the three-body level. The three expansion methods are not only useful for dealing with the complete potential, but they also provide separable approximations for the considered two-body potential.

Our study shows that a separable approximation works quite well for a shallow square-well potential, 
in which case there is only one $s$-wave two-body state that is bound ($a > 0$) or almost bound ($a < 0$). In this case, the Efimov states are true bound states. When we artificially move another two-body quasibound state closer to the three-body threshold, i.e., $E = 0$, the binding energies of the Efimov states shift, which shows that this quasibound state should be included as well in the description of the three-body system. In particular, we have found that strong $d$-wave interactions at positive scattering lengths have the effect of lowering the energy of the first excited Efimov state. This result is consistent with a recent study \cite{mestrom2017jila} in which strong $d$-wave interactions at positive scattering lengths are expected to be the cause of the noncrossing of the first excited Efimov state for a shallow Lennard-Jones potential.

The Efimov states are not only affected by two-body states in the continuum, but also by two-body bound states that are more deeply bound than the one that gives rise to the considered Efimov spectrum. Consequently, a separable approximation is insufficient for potentials that support multiple two-body bound states. For deep square-well potentials, the single-term EST approximation results in a three-body parameter $a_{-,0}$ that strongly deviates from the one calculated by the Weinberg expansion, which included many terms in the separable expansion of the $s$-wave component of the two-body $T$ matrix. This difference is caused by the fact that the single-term EST approximation of the full partial-wave component $t_0(p,p',z)$ only holds at small energies $z$, more precisely $|\bar{p}_z| \lesssim 0.5$, whereas the three-body equations involve $t_0(p,p',z)$ at all values of $z$ below the three-body energy $E$ for which solutions are sought. The deviation of $t_0(p,p',z)$ with its separable approximation at larger negative values of $z$ originates from the existence of strongly bound dimer states to which the particles can decay. 
When the potential becomes deeper, more two-body states with higher angular momentum quantum numbers $l$ become bound. Therefore, more partial-wave components of the off-shell $T$ matrix become important as well for the determination of the three-body parameter $a_{-,0}$. As a rule of thumb, we have found that whenever the potential is deep enough to (almost) support a two-body bound state with angular momentum quantum number $l$, then this partial-wave component should be included in the three-body calculation.

Furthermore, the separable approximation is insufficient to determine the three-body physics at large negative energies even at energies close to a particular two-body threshold where the off-shell $T$ matrix is highly separable. The failure of the separable approximation for negative three-body energies larger than the finite-range energy, i.e., $|E| \gtrsim \hbar^2 /(2 \mu R^2)$, has been attributed to the Green's function $G_0$ present in the Faddeev equations as discussed in \refSection{ssec:Results:shallow}. Therefore it is necessary to go beyond the separable approximation in this energy regime. Since this effect is related to the Faddeev equations themselves, we expect that this conclusion also holds for other potentials that describe atomic interactions more accurately. Therefore, it is likely that the nonseparability of the two-body interaction could affect the results of a recent study for $a>0$ in which a separable approximation was used to study atom-dimer scattering \cite{Greene2016softcoreVdW}. 

Even though a separable approximation based on the EST method gives reasonable results for the three-body parameter for certain classes of potentials \cite{naidon2014microscopic}, the square-well potential does not belong to any of those classes. Therefore it is interesting to analyze the features of the Efimov spectra corresponding to the square-well potential itself. Our results for this potential show that the three-body parameter $a_{-}$ is varying strongly for the lowest three potential resonances. However, our $s$-wave approach for deeper potentials up to 14 potential resonances suggests convergence for deep potentials. This change in the three-body parameter $a_{-}$ also affects the behavior of the Efimov states near the atom-dimer threshold. Even though the ground Efimov state does not merge with the atom-dimer threshold for the three lowest potential resonances, the first excited Efimov state only remains bound for the first potential resonance. Finally, we have found that additional trimer states are formed at atom-dimer threshold corresponding to the second and third $s$-wave dimer states as the potential depth is increased.

\section{acknowledgments}
The authors thank Jos\'e  P. D'Incao, Georg M. Bruun, Victor Colussi and Silvia Musolino for fruitful discussions. This research is financially supported by the Netherlands Organisation for Scientific Research (NWO) under Grant No. 680-47-623. 

\appendix
\section{The off-shell $T$ matrix of the square-well potential} \label{app:Toff_SqW}

Here we present the analytical expression for off-shell two-body $T$ matrix of the square-well potential. The method presented in Ref.~\cite{cheng1990matrix} can be used to find that the off-shell partial-wave components $t_l(p,p',z)$ of the square-well potential are given by
\begin{equation}\label{ToffmomentumSqW}
\begin{split}
t_l(p,p',z)&=\frac{R}{4\pi^2\mu\bar{p}\bar{p}'\hbar}\frac{\bar{q}^2-\bar{p}_z^2}{\bar{q}^2-\bar{p}^2}\\
&\left[\sigma(\bar{q};\bar{p},\bar{p}',\bar{p}_z)-\sigma(\bar{p};\bar{p},\bar{p}',\bar{p}_z)\right],
\end{split}
\end{equation}
where
\begin{equation}\label{sigma}
\begin{split}
\sigma(x;\bar{p},\bar{p}',\bar{p}_z)&=\left(\bar{p}_z^2-x^2\right)\\
&\frac{\bar{p}\bj_{l+1}(\bar{p})\hat{h}_l^{(1)}(\bar{p}_z)-\bar{p}_z\bj_l(\bar{p})\hat{h}_{l+1}^{(1)}(\bar{p}_z)}{x\bj_{l+1}(x)\hat{h}_l^{(1)}(\bar{p}_z)-\bar{p}_z\bj_l(x)\hat{h}_{l+1}^{(1)}(\bar{p}_z)}\\
&\cdot\frac{\bar{p}'\bj_{l+1}(\bar{p}')\bj_l(x)-x\bj_l(\bar{p}')\bj_{l+1}(x)}{\bar{p}'^2-x^2}.
\end{split}
\end{equation}
Here we have introduced the dimensionless momenta $\bar{p} = \frac{p R}{\hbar}$, $\bar{p}'=\frac{p' R}{\hbar}$, $\bar{p}_z=\frac{\sqrt{2 \mu z} R}{\hbar}$, $\bar{q} = \sqrt{\bar{q}_0^2 + \bar{p}_z^2}$ and $\bar{q}_0=\frac{\sqrt{2 \mu V_0} R}{\hbar}$. The Riccati-Bessel functions $\bj$, $\hat{n}_l$ and $\hat{h}_l^{(1)}$ are related to the usual spherical Bessel functions by $\bj(z) = z j_l(z)$, $\hat{n}_l(z)=-z n_l(z)$ and $\hat{h}_l^{(1)}(z) = \bj(z) - i \hat{n}_l(z)$.

The off-shell $s$-wave component $t_0(p,p',z)$ is related to the $s$-wave scattering length $a$ by
\begin{subequations}
\begin{align}
a &= 4 \pi^2 \mu \hbar \lim_{p,p',z\to 0}t_0(p,p',z) \label{eq:scattering_length_relation_t0}\\
&= R\left(1 - \frac{\tan(\bar{q}_0)}{\bar{q}_0} \right).\label{eq:SqW:scattering_length}
\end{align}
\end{subequations} 
The scattering length of the square-well potential thus diverges for $\bar{q}_0 = (2 N - 1) \pi/2$ where $N = 1, 2, 3, ...$ labels these potential resonances and counts the number of $s$-wave dimer states that are supported by this potential.

\section{Explicit form of the AGS integral equations} \label{app:AGS_integral_eq}

The three-body equations can be written out explicitly in the momentum-space representation. We expand the solution of these equations in a bispherical basis \cite{kharchenko1969separable,sitenko1991scattering} consisting of spherical harmonics and in terms of the form factors $g_{n l}(p,Z_q)$, so that we end up with a infinite set of one-dimensional integral equations.

In particular, for three-body recombination we derive from Eqs. (\ref{eq:Ua0_zero_E_limit}), (\ref{eq:Ubreve}) and (\ref{eq:K3zerovsU}) that $K_3(0)$ can be calculated from
\begin{equation}
\begin{aligned}
K_3(0) &= \frac{24 \pi m}{\hbar} (2 \pi \hbar)^6 \sum_{n_d,l_d} X_{n_d l_d}^2 \\
& q_d  \Bigg\lvert\sum_{n = 1}^{\infty}  \tau_{n, 0}\left(0\right) g_{n, 0}\left(0, 0 \right) A_{n_d l_d}^{n}(q_d, 0, 0)\Bigg\rvert^2,
\end{aligned}
\end{equation}
where the dimer states are labeled by the quantum numbers $n_d$ and $l_d$. For identical bosons the angular momentum quantum number $l_d$ is always even. Furthermore, the constant $X_{n l}$ relates the form factors of the expansion of $t_l(p,p',z)$ to the two-body bound state wave function in the momentum-space representation, $\langle \mathbf{p} | \varphi\rangle = \varphi_{n l}(p) Y_l^m(\hat{\mathbf{p}})$, according to
\begin{equation}
\varphi_{n l}(p) = X_{n l} \frac{g_{n l}(p,E_{2b,n l})}{E_{2b,n l} - \frac{p^2}{2 \mu}},
\end{equation}
where $E_{2b,n l}$ is the binding energy of the $n$th dimer state with angular momentum quantum number $l$. The factor $X_{n l}$ is thus simply a normalization constant, ensuring that $\langle\varphi | \varphi \rangle  = 1$. The momentum $q_d$ depends also on the indices $n_d$ and $l_d$ via $|E_{2b, n_d l_d}| = 3 q_d^2/(4 m) $. The amplitudes $A_{n l}^{n_i}(q,q_i,E)$ are calculated from
\begin{equation}\label{eq:amplitude_A}
\begin{aligned}
A_{n l}^{n_i}(q,&q_{i},E) = 2 U_{n l, n_i, 0}(q,q_{i},E) + 8 \pi \sum_{n',l'} \int_{0}^{\infty} \tau_{n' l'}\left(Z_{q'} \right)\\
& U_{n l,n' l'}(q,q',E) A_{n' l'}^{n_i}(q',q_{i},E) q'^2 \,dq',
\end{aligned}
\end{equation}
where the functions $U_{n l,n' l'}(q,q',E)$ are defined by
\begin{equation}
\begin{aligned}
U_{n l,n' l'}&(q,q',E)=\frac{1}{4 \pi}\Delta_l \Delta_{l'} \sqrt{2 l +1} \sqrt{2 l' +1}\\
&\int \frac{P_{l}(\mathbf{\hat{q}}\cdot\reallywidehat{\frac{1}{2} \mathbf{q} + \mathbf{q}'}) P_{l'}(\mathbf{\hat{q}'}\cdot\reallywidehat{\frac{1}{2} \mathbf{q}' + \mathbf{q}})}{\frac{1}{m}\left(q^2 + \mathbf{q}'\cdot \mathbf{q} + q'^2\right) - E }\\
& g_{n l}\left(\lvert\frac{1}{2} \mathbf{q} + \mathbf{q}'\rvert,Z_q\right) g_{n' l'}\left(\lvert\frac{1}{2} \mathbf{q}' + \mathbf{q}\rvert,Z_{q'}\right) \,d\mathbf{\hat{q}}'.
\end{aligned}
\end{equation}

For atom-dimer scattering in which case the considered dimer state is an $s$-wave bound state, we find from Eqs. (\ref{eq:AD_transition_amplitude}), (\ref{eq:aAD_related_to_U}), (\ref{eq:AD_Ugamma}) and (\ref{eq:AD_tildeUgamma}) that the $s$-wave atom-dimer scattering length can be calculated from
\begin{equation} \label{eq:AD_ScatL}
a_{ad} = -\frac{2}{3} \pi  m \hbar X_{n_i,0}^2 \lim_{q_i \to 0} A_{n_i,0}^{n_i}\left(q_i,q_i, E_{2b,i}+\frac{3}{4 m} q_i^2\right).
\end{equation}
Here the quantum number $n_i$ labels the considered $s$-wave dimer state whose bound state energy is $E_{2b,i}$. Remarkably, the atom-dimer scattering length is also related to the amplitudes $A_{n l}^{n_i}\left(q,q_{i},E\right)$ which we determine from \refEquation{eq:amplitude_A} for small values of $q_{i}$, namely $q_{i} a/\hbar = 10^{-5}$, so that the kinetic energy at which the atom and dimer scatter is much smaller than the binding energy of the dimer. The details of deriving \refEquation{eq:AD_ScatL} can be found in Ref.~\cite{kharchenko1969separable,sitenko1991scattering}.

The set of equations given by \refEquation{eq:amplitude_A} involves singularities caused by the factor $\tau_{n l}(Z_{q'})$. These poles are treated by splitting the integral into a principal value integral along the real axis and a complex part proportional to the residue of the integrand. In the special case that a dimer in the two-body ground state  scatters with a free particle at zero energy, there is only one singularity for which only the principal value part of the singular integral matters because the residue is zero in the limit $q_{i} \rightarrow 0$. Therefore the atom-dimer scattering length is real in this particular case.

\newpage
\section{Additional data} \label{app:tables}

This section contains some tables supporting our conclusions of \refSection{sec:Results}.

\begin{table}[H]
  \centering
  \caption{Values of the three-body parameters $\bar{a}_{-,n}$ and $\bar{\kappa}_n^*$  corresponding to the first potential resonance of the square-well potential using different methods and different number of terms to expand $t_0(p,p',z)$ and $t_2(p,p',z)$. 
Method I refers to the spectral representation, method II refers to the Weinberg expansion, and method III refers to the single-term EST approximation as discussed in \refSection{ssec:EST}.
  Method I$^{*}$ refers to method I in which the $d$-wave resonance is artificially shifted from $\bar{a} = 0$ to $\bar{a} = 1$.}
  \label{tab:3BP_SqW_PotRes1}
  \begin{ruledtabular}
    \begin{tabular}{ccccccc}
    \toprule
    method & $N_s$ & $N_d$  & $\bar{a}_{-,0}$ & $\bar{a}_{-,1}$ & $\bar{\kappa}_0^*$ & $\bar{\kappa}_1^*$ \\
    \hline
    I & 1 & 0 & -3.102 & -55.23 & 0.6647 & $2.831\cdot10^{-2}$\\
    I & 2 & 0 & -3.092 & -54.96 & 0.6654 & $2.845\cdot10^{-2}$\\
    I & 3 & 0 & -3.091 & -54.94 & 0.6655 & $2.846\cdot10^{-2}$\\
    I & 4 & 0 & -3.091 & -54.93 & 0.6655 & $2.846\cdot10^{-2}$\\
    I & 5 & 0 & -3.090 & -54.93 & 0.6655 & $2.846\cdot10^{-2}$\\
    I & 3 & 1 & -3.088 & -54.91 & 0.6661 & $2.848\cdot10^{-2}$\\
    I & 3 & 2 & -3.088 & -54.91 & 0.6662 & $2.848\cdot10^{-2}$\\
    I & 3 & 3 & -3.088 & -54.91 & 0.6662 & $2.848\cdot10^{-2}$\\
    I$^{*}$ & 3 & 3 & -3.072 & -54.78 & 0.6689 & $2.854\cdot10^{-2}$\\
    II & 1 & 0 & -3.163 & -55.79 & 0.6536 & $2.804\cdot10^{-2}$\\
    II & 2 & 0 & -3.104 & -55.09 & 0.6631 & $2.838\cdot10^{-2}$\\
    II & 3 & 0 & -3.095 & -54.98 & 0.6647 & $2.844\cdot10^{-2}$\\
    II & 4 & 0 & -3.092 & -54.95 & 0.6652 & $2.845\cdot10^{-2}$\\
    II & 5 & 0 & -3.091 & -54.94 & 0.6653 & $2.846\cdot10^{-2}$\\
    II & 10 & 0 & -3.090 & -54.93 & 0.6655 & $2.847\cdot10^{-2}$\\
    III & 1 & 0 & -3.106 & -55.51 & 0.6610 & $2.815\cdot10^{-2}$\\
    \bottomrule
    \end{tabular}\\
    \end{ruledtabular}
\end{table}


\begin{table}[H]
  \centering
  \caption{Values for the three-body parameters $\bar{a}_{-,n}$ and $\bar{\kappa}_n^*$  corresponding to the $N$th potential resonance of the square-well potential using the single-term EST approximation for $t_0(p,p',z)$. In \refAppendixSection{app:Toff_SqW} the depth of the square well is related to the number $N$.}
  \label{tab:3BP_SqW_methodEST}
  \begin{ruledtabular}
    \begin{tabular}{ccccc}
    \toprule
    $N$  & $\bar{a}_{-,0}$ & $\bar{a}_{-,1}$ & $\bar{\kappa}_0^*$ & $\bar{\kappa}_1^*$ \\
    \hline
    1 & -3.106 & -55.51 & 0.6610 & $2.815\cdot10^{-2}$\\
    2 & -12.58 & -260.8 & 0.1332 & $5.835\cdot10^{-3}$\\
    3 & -13.80 & -290.3 & 0.1193 & $5.235\cdot10^{-3}$\\
    4 & -13.12 & -274.5 & 0.1262 & $5.539\cdot10^{-3}$\\
    5 & -13.38 & -280.7 & 0.1233 & $5.415\cdot10^{-3}$\\
    6 & -13.54 & -284.2 & 0.1219 & $5.348\cdot10^{-3}$\\
    7 & -12.86 & -269.0 & 0.1288 & $5.654\cdot10^{-3}$\\
    8 & -13.03 & -272.8 & 0.1270 & $5.573\cdot10^{-3}$\\
    9 & -13.08 & -273.9 & 0.1265 & $5.551\cdot10^{-3}$\\
    10 & -13.13 & -275.0 & 0.1259 & $5.528\cdot10^{-3}$\\
    \vdots & \vdots & \vdots & \vdots & \vdots \\
    49 & -13.23 & -277.3 & 0.1249 & $5.482\cdot10^{-3}$\\
    50 & -13.23 & -277.3 & 0.1249 & $5.482\cdot10^{-3}$\\
    \vdots & \vdots & \vdots & \vdots & \vdots \\
    $\infty$ & -13.24 & -277.4 & 0.1249 & $5.481\cdot10^{-3}$\\
    \bottomrule
    \end{tabular}\\
    \end{ruledtabular}
\end{table}

\begin{table}[H]
  \centering
  \caption{Values for the three-body parameters $\bar{a}_{-,0}$ corresponding to the $N$th potential resonance of the square-well potential using the Weinberg expansion to expand the $s$-wave component $t_0(p,p',z)$. The other partial-wave components of the off-shell $T$ matrix are not included in these three-body calculations, so that the values for $\bar{a}_{-,0}$ should not be regarded as the true three-body parameter of the square-well potential as shown by Table \ref{tab:3BP_SqW_methodVG0_l_total}. Two methods (A and B) are considered. The relevant eigenvalue $\varepsilon$ of the three-body kernel is also given in this table for method A. The sign of the imaginary part depends on the integration contour ($\lim_{\epsilon \rightarrow 0} E \pm i \epsilon$). The loss parameter $\eta_{*}$ is also listed for method B.}
  \label{tab:3BP_SqW_methodVG0_l_0}
  \begin{ruledtabular}
    \begin{tabular}{ccccc}
    \toprule
    $N$  & $\bar{a}_{-,0}$ & $\varepsilon-1$ & $\bar{a}_{-,0}$ & $\eta_{*}$ \\
     & (A) & (A) & (B) & (B)\\
    \hline
    1 & -3.09 & 0 & - & -\\
    2 & -16.8 &  $\pm$ 0.014 i & -17.0 & 0.050\\
    3 & -11.0 &  $\pm$ 0.074 i & -19.1 & 0.134\\
    4 & -21.5 &  $\pm$ 0.031 i & -22.8 & 0.108\\ 
    5 & -24.8 &  $\pm$ 0.042 i & -25.0 & 0.121\\ 
    6 & -23.1 &  $\pm$ 0.019 i & -23.9 & 0.078\\ 
    7 & -24.8 &  $\pm$ 0.018 i & -25.2 & 0.061\\ 
    8 & -22.3 &  $\pm$ 0.008 i & -22.3 & 0.037\\ 
    9 & -24.1 &  $\pm$ 0.010 i & -24.3 & 0.045\\ 
    10 & -24.0 &  $\pm$ 0.018 i & -24.6 & 0.101\\ 
    11 & -23.8 &  $\pm$ 0.011 i & -23.8 & 0.060\\ 
    12 & -23.8 &  $\pm$ 0.013 i & -24.6 & 0.083\\ 
    13 & -25.3 &  $\pm$ 0.017 i & -25.6 & 0.081\\ 
    14 & -24.2 &  $\pm$ 0.014 i & -25.0 & 0.082\\ 
    \bottomrule
    \end{tabular}\\
    \end{ruledtabular}
\end{table}

\begin{table}[H]
  \centering
  \caption{Values for the three-body parameters $\bar{a}_{-,0}$ corresponding to the $N$th potential resonance of the square-well potential as a function of the partial-wave components $t_l(p,p',z)$ that are included in the calculation. The Weinberg expansion is used to expand the functions $t_l(p,p',z)$. Two methods (A and B) are considered. The relevant eigenvalue $\varepsilon$ of the three-body kernel is also given in this table for method A. The sign of the imaginary part depends on the integration contour ($\lim_{\epsilon \rightarrow 0} E \pm i \epsilon$). The loss parameter $\eta_{*}$ is also listed for method B.}
  \label{tab:3BP_SqW_methodVG0_l_total}
  \begin{ruledtabular}
    \begin{tabular}{clcccc}
	\toprule
    $N$ & $l$ & $\bar{a}_{-,0}$ & $\varepsilon-1$ & $\bar{a}_{-,0}$ & $\eta_{*}$ \\
     & & (A) & (A) & (B) & (B)\\
    \hline    
	2 & $[0]$ & -16.8 &  $\pm$ 0.014 i & -17.0 & 0.050\\
    2 & $[0, 2]$ & -18.1 &  $\pm$ 0.010 i & -18.2 & 0.047\\
    2 & $[0, 2, 4]$ & -17.3 &  $\pm$ 0.012 i & -17.4 & 0.059\\ 
    2 & $[0, 2, 4, 6]$ & -17.3 &  $\pm$ 0.012 i & -17.4 & 0.060\\ 
    2 & $[0, 2, 4, 6, 8]$ & -17.3 &  $\pm$ 0.012 i & -17.4 & 0.060\\  
    3 & $[0]$ & -11.0 &  $\pm$ 0.074 i & -19.1 & 0.134\\
    3 & $[0, 2]$ & -23.0 &  $\pm$ 0.031 i & -22.2 & 0.103\\
    3 & $[0, 2, 4]$ & -27.7 &  $\pm$ 0.015 i & -27.3 & 0.063\\
    3 & $[0, 2, 4, 6]$ & -25.9 &  $\pm$ 0.004 i & -25.8 & 0.021\\
    3 & $[0, 2, 4, 6, 8]$ & -25.7  &  $\pm$ 0.004 i & -25.7 & 0.021\\
    \bottomrule
    \end{tabular}\\
    \end{ruledtabular}
\end{table}

\bibliographystyle{apsrev}
\bibliography{Bibliography}

\begin{thebibliography}{79}
\bibitem[{not()}]{noteCalcScatLength}
\bibinfo{note}{In all cases scattering lengths are calculated from the
  scattering length of the square-well potential given by
  \refEquation{eq:SqW:scattering_length}, in order to ensure the proper
  comparison between the methods, even when the approximated potentials have a
  slightly different scattering length.}
\bibitem[{rep()}]{repeated3bodyCalc}
\bibinfo{note}{We note that the calculation of the Efimov spectra involving
  only $s$-wave interactions has been repeated by using the Weinberg expansion
  (method II), leading to the same Efimov spectra as the spectral
  representation (method I). Those calculations also show that the first
  excited Efimov state of the shallow square-well potential stays below the
  atom-dimer threshold when we use the single-term approximation of the
  Weinberg expansion to approximate $t_0(p,p',z)$.}
\end{thebibliography}

%

\end{document}